\documentstyle[12pt,epsf]{article}
\newlength{\pubnumber} 

\catcode`\@=11
\@addtoreset{equation}{section}

\def\section{\@startsection{section}{1}{\z@}{3.5ex plus 1ex minus .2ex}
 {2.3ex plus .2ex}{\large\bf}}
\def\subsection{\@startsection{subsection}{2}{\z@}{2.3ex plus .2ex}
 {2.3ex plus .2ex}{\bf}}

\def\tr{{\rm tr} }

\def\PS{PS  }
\def\FSU{FSU5  }
\def\LRS{LRS  }
\def\SLM{SLM  }

\def\PSc{PS}
\def\FSUc{FSU5}

\def\SLMc{SLM}

\input epsf

\begin{document}

\begin{titlepage}
\samepage{
\setcounter{page}{1}
\rightline{BU--HEPP--02/13}
\rightline{CASPER--02/05}
\rightline{OUTP-03--01P}

\rightline{\tt hep-ph/0301037}
\rightline{December 2002}
\vfill
\begin{center}
 {\Large \bf NAHE--Based String Models \\
With $SU(4)\times SU(2)\times U(1)$ $SO(10)$ Subgroup \\}
\vfill
\vfill {\large
	Gerald B.
Cleaver$^{1}$\footnote{gerald\underline{\phantom{B}}cleaver@baylor.edu},
        Alon E. Faraggi$^{2}$\footnote{faraggi@thphys.ox.ac.uk}
	and
	Sander Nooij$^{2}$\footnote{semnooij@thphys.ox.ac.uk}}\\
\vspace{.12in}
{\it $^{1}$ CASPER, Department of Physics,
            Baylor University,\\
            Waco, TX, 76798-7316, USA\\
            and\\
             Astro Particle Physics Group,
            Houston Advanced Research Center (HARC),\\
            The Mitchell Campus,
            Woodlands, TX 77381, USA\\}
\vspace{.05in}
{\it $^{2}$ Theoretical Physics Department, University of Oxford,
            Oxford, OX1 3NP, UK\\}
\vspace{.025in}
\end{center}
\vfill
\begin{abstract}
The orbifold GUT doublet--triplet splitting mechanism was
discussed in 1994 in the framework of the NAHE--based free
fermionic models in which the $SO(10)$ GUT symmetry is broken
to $SO(6)\times SO(4)$, $SU(3)\times SU(2)\times U(1)^2$,
or $SU(3)\times U(1)\times SU(2)^2$.
In this paper we study NAHE--based free fermionic models in
which the $SO(10)$ symmetry is broken at the string level
to $SU(4)\times SU(2)\times U(1)$. In addition to the
doublet--triplet splitting this case also has the
advantage of inducing the doublet--doublet splitting
already at the string level. We demonstrate, however, that
NAHE--based models with $SU(4)\times SU(2)\times U(1)$
$SO(10)$ subgroup are not viable. We show that, similarly to the
\LRS models, and in contrast to the FSU5, PS and SLM models,
the SU421 case gives rise to models without an anomalous $U(1)$
symmetry, and discuss the different cases in terms of their
$N=4$ origins.

\end{abstract}
\smallskip}
\end{titlepage}

\setcounter{footnote}{0}

\def\beq{\begin{equation}}
\def\eeq{\end{equation}}
\def\beqn{\begin{eqnarray}}
\def\eeqn{\end{eqnarray}}

\def\no{\noindent }
\def\nolabel{\nonumber }
\def\ie{{\it i.e.}}
\def\eg{{\it e.g.}}
\def\half{{\textstyle{1\over 2}}}
\def\third{{\textstyle {1\over3}}}
\def\quarter{{\textstyle {1\over4}}}
\def\sixth{{\textstyle {1\over6}}}
\def\m{{\tt -}}
\def\p{{\tt +}}

\def\Tr{{\rm Tr}\, }
\def\tr{{\rm tr}\, }

\def\slash#1{#1\hskip-6pt/\hskip6pt}
\def\slk{\slash{k}}
\def\GeV{\,{\rm GeV}}
\def\TeV{\,{\rm TeV}}
\def\y{\,{\rm y}}
\def\SM{Standard--Model }
\def\SUSY{supersymmetry }
\def\SSSM{supersymmetric standard model}
\def\vev#1{\left\langle #1\right\rangle}
\def\l{\langle}
\def\r{\rangle}
\def\o#1{\frac{1}{#1}}

\def\Htw{{\tilde H}}
\def\chibar{{\overline{\chi}}}
\def\qbar{{\overline{q}}}
\def\ibar{{\overline{\imath}}}
\def\jbar{{\overline{\jmath}}}
\def\Hbar{{\overline{H}}}
\def\Qbar{{\overline{Q}}}
\def\abar{{\overline{a}}}
\def\alphabar{{\overline{\alpha}}}
\def\betabar{{\overline{\beta}}}
\def\tautwo{{ \tau_2 }}
\def\thetatwo{{ \vartheta_2 }}
\def\thetathree{{ \vartheta_3 }}
\def\thetafour{{ \vartheta_4 }}
\def\ttwo{{\vartheta_2}}
\def\tthree{{\vartheta_3}}
\def\tfour{{\vartheta_4}}
\def\ti{{\vartheta_i}}
\def\tj{{\vartheta_j}}
\def\tk{{\vartheta_k}}
\def\calF{{\cal F}}
\def\smallmatrix#1#2#3#4{{ {{#1}~{#2}\choose{#3}~{#4}} }}
\def\ab{{\alpha\beta}}
\def\Minv{{ (M^{-1}_\ab)_{ij} }}
\def\bone{{\bf 1}}
\def\bo{{\bf 1}}
\def\ii{{(i)}}
\def\V{{\bf V}}
\def\N{{\bf N}}

\def\bfb{{\bf b}}
\def\bfS{{\bf S}}
\def\bfX{{\bf X}}
\def\bfI{{\bf I}}
\def\ma{{\mathbf a}}
\def\mb{{\mathbf b}}
\def\mS{{\mathbf S}}
\def\mX{{\mathbf X}}
\def\mI{{\mathbf I}}
\def\malpha{{\mathbf \alpha}}
\def\mbeta{{\mathbf \beta}}
\def\mgamma{{\mathbf \gamma}}
\def\mzeta{{\mathbf \zeta}}
\def\mxi{{\mathbf \xi}}

\def\t#1#2{{ \Theta\left\lbrack \matrix{ {#1}\cr {#2}\cr }\right\rbrack }}
\def\C#1#2{{ C\left\lbrack \matrix{ {#1}\cr {#2}\cr }\right\rbrack }}
\def\tp#1#2{{ \Theta'\left\lbrack \matrix{ {#1}\cr {#2}\cr }\right\rbrack }}
\def\tpp#1#2{{ \Theta''\left\lbrack \matrix{ {#1}\cr {#2}\cr }\right\rbrack }}
\def\l{\langle}
\def\r{\rangle}

\def\cL#1#2{{\cal L}^{#1}_{#2}}
\def\x#1{\phi_{#1}}
\def\bx#1{{\bar{\phi}}_{#1}}

\def\cl#1{{\cal L}_{#1}}
\def\bcl#1{\bar{\cal L}_{#1}}

\def\bt{{\bar 3}}
\def\h#1{h_{#1}}
\def\Q#1{Q_{#1}}
\def\L#1{L_{#1}}

\def\N#1{N_{#1}}
\def\bN#1{{\bar{N}}_{#1}}

\def\S#1{S_{#1}}
\def\Ss#1#2{S_{#1}^{#2}}
\def\bS#1{{\bar{S}}_{#1}}
\def\bSs#1#2{{\bar{S}}_{#1}^{#2}}

\def\D#1{D_{#1}}
\def\Ds#1#2{D_{#1}^{#2}}
\def\bD#1{{\bar{D}}_{#1}}
\def\bDs#1#2{{\bar{D}}_{#1}^{#2}}

\def\p#1{\phi_{#1}}
\def\bp#1{{\bar{\phi}}_{#1}}

\def\P#1{\Phi_{#1}}
\def\bP#1{{\bar{\Phi}}_{#1}}
\def\X#1{\Phi_{#1}}
\def\bX#1{{\bar{\Phi}}_{#1}}
\def\Ps#1#2{\Phi_{#1}^{#2}}
\def\bPs#1#2{{\bar{\Phi}}_{#1}^{#2}}
\def\ps#1#2{\phi_{#1}^{#2}}
\def\bps#1#2{{\bar{\phi}}_{#1}^{#2}}
\def\php{\phantom{+}}

\def\H#1{H_{#1}}
\def\bH#1{{\bar{H}}_{#1}}

\def\xH#1#2{H^{#1}_{#2}}
\def\bxH#1#2{{\bar{H}}^{#1}_{#2}}


\def\inbar{\,\vrule height1.5ex width.4pt depth0pt}

\def\IC{\relax\hbox{$\inbar\kern-.3em{\rm C}$}}
\def\IQ{\relax\hbox{$\inbar\kern-.3em{\rm Q}$}}
\def\IR{\relax{\rm I\kern-.18em R}}
 \font\cmss=cmss10 \font\cmsss=cmss10 at 7pt
\def\IZ{\relax\ifmmode\mathchoice
 {\hbox{\cmss Z\kern-.4em Z}}{\hbox{\cmss Z\kern-.4em Z}}
 {\lower.9pt\hbox{\cmsss Z\kern-.4em Z}}
 {\lower1.2pt\hbox{\cmsss Z\kern-.4em Z}}\else{\cmss Z\kern-.4em Z}\fi}

\def\AEF{A.E. Faraggi}
\def\NPB#1#2#3{{\it Nucl.\ Phys.}\/ {\bf B#1} (#2) #3}
\def\PLB#1#2#3{{\it Phys.\ Lett.}\/ {\bf B#1} (#2) #3}
\def\PRD#1#2#3{{\it Phys.\ Rev.}\/ {\bf D#1} (#2) #3}
\def\PRL#1#2#3{{\it Phys.\ Rev.\ Lett.}\/ {\bf #1} (#2) #3}
\def\PRT#1#2#3{{\it Phys.\ Rep.}\/ {\bf#1} (#2) #3}
\def\MODA#1#2#3{{\it Mod.\ Phys.\ Lett.}\/ {\bf A#1} (#2) #3}
\def\IJMP#1#2#3{{\it Int.\ J.\ Mod.\ Phys.}\/ {\bf A#1} (#2) #3}
\def\nuvc#1#2#3{{\it Nuovo Cimento}\/ {\bf #1A} (#2) #3}
\def\RPP#1#2#3{{\it Rept.\ Prog.\ Phys.}\/ {\bf #1} (#2) #3}
\def\etal{{\it et al\/}}

\hyphenation{su-per-sym-met-ric non-su-per-sym-met-ric}
\hyphenation{space-time-super-sym-met-ric}
\hyphenation{mod-u-lar mod-u-lar--in-var-i-ant}


\setcounter{footnote}{0}
\section{Introduction}
Grand unified extensions of the Standard Particle Model are well supported
by the phenomenological Standard Model data. This is reflected by
the particle charges, by the extrapolation of the gauge,
and heavy generation mass, parameters, and by the suppression of
proton decay and neutrino masses. The most appealing GUT embedding
is achieved with an $SO(10)$ gauge group in which each of
Standard Model generations is embedded in a single 16 $SO(10)$
representation. However, it is clear that
a more complete understanding of the Standard Model parameters
necessitates the incorporation of gravity in the unification program.
This conclusion follows from the fact that flavor and mass
do not originate from Grand Unified Theories. These origins
must then be sought in the scale above the GUT scale, namely the
Planck scale, and therefore in a theory of quantum gravity.

String theory is the unique contemporary framework that enables
detailed studies of the gauge--gravitational unification.
Given the success of the Standard Model and Grand Unified Theories
it is natural to seek string models that preserve the GUT
embedding of the Standard Model spectrum. The GUT embedding
may be realized at the string level and is not necessarily
present in the effective low energy field
theory. Most compelling in this context are models that
preserve the $SO(10)$ embedding of the Standard Model spectrum.

Following the string duality developments we now believe that
the string theories are perturbative limits of a still unknown
more fundamental theory. In this respect, we should think of
the string theories as probes of the properties of the more
basic structures. In this view, different string limits may
be advantageous in probing different characteristics of the
underlying theory. So, for example, the issue of supersymmetry
breaking and dilaton stabilization may be better addressed
in the type I limit. On the other hand the limit that enables
the $SO(10)$ embedding of the Standard Model spectrum is the
heterotic--string limit as it is the only limit that produces
the chiral 16 representation of $SO(10)$ in the perturbative
massless spectrum.
In addition to the desirable GUT embedding of the Standard Model
spectrum one must also impose that the perturbative massless
spectrum contains three chiral generations.

A class of three generation heterotic--string compactifications
that preserves the $SO(10)$ embedding of the Standard Model spectrum
was constructed in the so--called free fermionic formulation.
The existence of models in this class which produce solely
the spectrum of the Minimal Supersymmetric Standard Model
in the observable massless sector was further demonstrated \cite{cfn1}.
The free fermionic models correspond to $Z_2\times Z_2$
orbifold compactification at a maximally symmetric point
in the Narain moduli space and additional Wilson lines.
The realistic features of the free fermionic models are rooted
in the underlying $Z_2\times Z_2$ orbifold structure,
and it is of course of much interest to examine whether
there exists a dynamical reason why this class of compactifications
would be selected.

The $SO(10)$ symmetry is broken to one of its subgroups
by the free fermion boundary condition basis vectors,
that are equivalent to Wilson lines in the orbifold
formulation. To date $SO(10)$ subgroups that have been studied
include the flipped $SU(5)$ (FSU5) \cite{fsu5};
the standard--like models (SLM) \cite{fny,slm,cfn1};
the Pati--Salam models (PS) \cite{alr};
and the left--right symmetric models (LRS) \cite{cfs}.
Many of the phenomenological properties of the models
naturally relate to the $SO(10)$ subgroup which remains
unbroken. For example, the stringy doublet--triplet mechanism
of ref \cite{ps}
operates in the SLM, PS, and LRS models but not in the FSU5
models\footnote{Similar doublet--triplet splitting mechanisms
were also discussed in the context of other Calabi--Yau and
orbifold constructions\cite{otherdts}. The fermionic models afforded
the understanding of the splitting in terms of the GSO projections and
the world--sheet fermionic boundary conditions, that correspond
to action on the internal dimensions in the corresponding
bosonic language \cite{ps}. Alternatively, an attractive field
theory mechanism is obtained in the flipped $SU(5)$ \cite{fsu5dts}.}.
Similarly, the top--bottom mass splitting mechanism
of ref. \cite{yukawa} operates in the SLM and FSU5 models,
but not in the PS or LRS models. Another important example
is the relation between the presence of an anomalous
$U(1)$ and the choice of the unbroken $SO(10)$ subgroup.
In the case of the FSU5, SLM and PS one always finds
an anomalous $U(1)$, whereas LRS models sometimes
yield three generation models that are free from any
Abelian and non--Abelian anomalies. This aspect is particularly
intriguing as it relates to the issue of supersymmetry breaking
in the string models. Specifically, it was demonstrated in
ref. \cite{ccf} that in LRS models that do contain an
anomalous $U(1)$ there exist only non--Abelian flat directions,
whereas Abelian flat--directions do no exist.

To date, the class of free fermionic models with unbroken
$SU(4)\times SU(2) \times U(1)$ (SU421)
$SO(10)$ subgroup has not been analyzed
in the literature. In this paper we undertake this analysis,
and by that complete the analysis of all the possible unbroken
$SO(10)$ subgroups. We demonstrate that in fact, with the NAHE set basis
vectors \cite{nahe}, this choice cannot produce realistic spectrum.
Specifically, we demonstrate that this choice of $SO(10)$ symmetry
breaking pattern necessarily results in incomplete $SO(10)$ multiplets 
and hence does not produce the Standard Model chiral matter.
Additionally, we show that, similarly to the LRS models, SU421 models
can produce anomaly free, or anomalous, $U(1)$ models, depending
on the initial $N=4$ vacuum. The existence or absence of an anomalous
$U(1)$ hinges on the important issue of supersymmetry breaking.
In ref. \cite{ccf} we argued that the absence of Abelian flat directions
in anomalous LRS models indicates that supersymmetry is broken hierarchically
in the models by hidden sector gaugino and matter condensation.
However, here we find that, unlike the LRS models,
singlet flat directions also exist in the SU421 models that
do contain an anomalous $U(1)$.

\setcounter{footnote}{0}
\section{The supersymmetric $SU(4)\times SU(2)\times U(1)$
model}\label{leftrightft}

In this section we briefly summarize the field theory
structure of the type of models that we aim to construct from
string theory in this paper. The observable sector gauge symmetry
we seek is
$SU(4)_C\times SU(2)_L\times U(1)_R$.
Such models are reminiscent of the \PS
type string models, but differ from them by the fact that the $SU(2)_R$
gauge group is broken to $U(1)_{R}$ already at the
string level. Similar to the \PS models, the
$SU(4)_C\times SU(2)_L \times U(1)_R$
models possess the $SO(10)$ embedding.
The quarks and leptons are accommodated in the following
representations:
\beqn
F_L^{i} &=& (      4 ,2,~~0)     ~~=~
	    (3 ,2, ~~{1\over3},~~~0) + (1,2, -{1},~~~0)
       ~~=~  {u\choose d}^i+{\nu\choose e}^i,\label{SU421fl}\\
U_R^{i} &=& ({\bar 4},1,-{1\over2})    ~=~
	    ({\bar 3},1,-{1\over3},-{1\over2}) + (1,1,+{1},-{1\over2})
       ~~=~{u^{c}}^i+{N^{c}}^i,\label{SU421ur}\\
D_R^{i} &=& ({\bar 4},1,+{1\over2})    ~=~
	    ({\bar 3},1,-{1\over3},+{1\over2}) + (1,2,+{1},+{1\over2})
       ~~=~{d^{c}}^i+{e^{c}}^i,
\label{SU421dr}
\eeqn
where the first and second equalities display the decomposition under
$SU(4)_C\times SU(2)_L\times U(1)_R$ and under
$SU(3)_C\times SU(2)_L\times U(1)_{B-L}\times U(1)_R$, respectively,
and the weak--hypercharge is given by
\beq
U(1)_Y={1\over2}U(1)_{B-L}+U(1)_R.
\label{weakhypecharge}
\eeq
Hence, $F_L$ produces the quark and lepton electroweak doublets
whereas, $U_R$ and $D_R$ produces the right--handed electroweak
singlets. The two low energy supersymmetric Higgs superfields
associated with the Minimal Supersymmetric Standard Model,
$h^d$ and $h^u$, are given by,
\beqn
h^d     &=& (      1 ,2,-{1\over2}),  \\
h^u     &=& (      1 ,2,+{1\over2}).
\label{SU421mssmhigss}
\eeqn
The heavy Higgs multiplets that break $SU(4)_C\times U(1)_{R}$
to the Standard Model group factors $SU(3)\times U(1)_Y$
are given by the fields
\beqn
     {H} &=& ({\bar 4},1,+1)\\
{\bar H} &=& ({     4},1,-1)
\label{SU421Higgs}
\eeqn
The  SU421 string models can also contain Higgs fields that transform as
$$(6,1,0)= ({3},1,{1\over3},0)+ ({\bar 3},1,-{1\over3},0),$$
that originates from the vectorial {\bf 10} representation of $SO(10)$.
These color triplets mediate proton decay through dimension five operators,
and consequently must be sufficiently heavy to insure agreement with
the proton lifetime. An important advantage of the SU421 symmetry
breaking pattern, with
$SO(10)\rightarrow SO(6)\times SO(4)\rightarrow SU(4)\times SU(2)\times U(1)$
at the string construction level,
is that these color triplets may be projected out by the GSO
projections \cite{ps}, and therefore need not be present in the
low energy spectrum. In principle, the string doublet--triplet splitting
mechanism operates in all the models that include the symmetry
breaking pattern $SO(10)\rightarrow SO(6)\times SO(4)$.
In the \PS models, however,
the Higgs representations that induce $SU(4)\times SU(2)_R\rightarrow
SU(3)_C\times U(1)_{Y}$ contain the Higgs triplet representations
with the quantum numbers of eq. (\ref{SU421dr}),
that may mediate rapid proton decay through dimension five operators.
In the supersymmetric \PS models the color triplets
in the vectorial representation $(6,1,1)$ are used to give
large mass to the Higgs color triplets, by the superpotential
terms $\lambda_2HHD+\lambda_3 {\bar H}{\bar H}{\bar D}$, when the fields
$H$ and ${\bar H}$ develop a large VEV of the order of the GUT scale.
On the other hand, in the SU421 models the Higgs representations of
eq. (\ref{SU421Higgs}) do not contain the fields with the
quantum numbers of eq. (\ref{SU421dr})
Therefore, the string doublet--triplet splitting mechanism is useful
only in models with $SU(3)_C\times SU(2)_L\times U(1)^2$,
$SU(4)_C\times SU(2)_L\times U(1)_R$, or
$SU(3)_C\times SU(2)_L\times SU(2)_R \times U(1)_{B-L}$,
as the $SO(10)$ subgroup which remains unbroken by the GSO projections.

Another important advantage of the SU421 string models is
with respect to the electroweak Higgs doublet representations.
In the left--right symmetric models, {\it i.e.} the \PS and \LRS
models in which $SU(2)_L$ and $SU(2)_R$ both remain unbroken at the string
scale, up--quark and down--quark masses both arise from the coupling
to a Higgs bi--doublet. This introduces the danger of inducing Flavor
Changing Neutral Currents (FCNC) at an unacceptable rate \cite{lrsfcnc}.
A possible  solution is to use two bi--doublet Higgs representations,
one of which is used to give masses to the up--type quarks, while
the second is used to give masses to the down--type quarks.
This, however, introduces a bi--doublet splitting problem.
Namely, we must insure that one Higgs multiplet
remains light to give mass to the up-- or down--type quarks, while
the second Higgs multiplet in the respective bi--doublet
becomes sufficiently heavy so as to avoid problems with FCNC.
Arguably, this can be achieved in a field theory setting. However,
the bi--doublet splitting mechanisms that have been discussed
in the literature \cite{bidoublet}
utilize $SU(2)$ triplet representations that are,
in general, not present in the free fermionic string models.
Therefore, whether
or not bi--doublet splitting can be achieved in the left--right symmetric
string models is an open question. By contrast, in the SU421 models
$SU(2)_R$ is broken at the string level. Consequently, the bi--doublet Higgs
is split already at the string level. The up and down quarks therefore
couple to separate doublet Higgs multiplets, and the problem with
FCNC is evaded.

The SU421 models should also contain four additional
singlet fields $\phi_0$ and $\phi_{i=1,2,3}$. $\phi_0$ acquires
a VEV of the order of the electroweak scale which induces the
electroweak Higgs doublet mixing, while $\phi_i$ are used
to construct an extended see--saw mechanism that generates
light Majorana masses for the left--handed neutrinos.
The tree level superpotential of the model is given by:
\beq
W=\lambda^1_{ij}F_L^iU_R^j{\bar h}+\lambda^2_{ij}F_L^iD_R^jh+
\lambda^3_{ij}U_R^i{\bar H}\phi^j+
\lambda^4{\bar h}h\phi^0+\lambda_5\Phi^3,
\label{superpot}
\eeq
where $\Phi=\{\phi^i,\phi^0\}$. The superpotential in
eq.~(\ref{superpot}) leads to the neutrino mass matrix
\beq
{\left(\matrix{
                 0 & m_u^{ij}        &  0           \cr
         m_u^{ji}  &   0             &  \l{\bar H}\r\cr
                 0 &   \l{\bar H}\r  &  \l\phi_0\r}\right)},
\label{seesaw}
\eeq
whose diagonalization gives three light neutrinos with masses
of the order $\l\phi_0\r(m_u^{ij}/\l{\bar H}\r)^2$ and gives
heavy mass, of order $\l{\bar H}\r$, to the right--handed neutrinos.

Below the scale of $SU(4)_C\times U(1)_R$
breaking the SU421 models
should reproduce the spectrum and couplings of the MSSM.
As we have seen SU421 string models offer important advantages with
respect to the doublet--triplet and bi--doublet splitting problems.
A field theory analysis of these models is therefore of
further interest.
As our interest
here is primarily in the string construction of SU421
models we do not enter into further field theory details.
We emphasize that our intent here is not to construct a fully
realistic SU421 model, but merely to study the structure of
the NAHE--based
free fermionic string models with this choice of the $SO(10)$
subgroup. In this respect we note that the bi--doublet
splitting problem introduces further motivation for
the choice of $SU(3)\times SU(2)\times U(1)^2$ or
$SU(4)\times SU(2)\times U(1)$ as
the $SO(10)$ subgroup that remain unbroken after
application of the string GSO projections.
Thus, while the doublet--triplet splitting problem
does not distinguish between the \PS string model ($SO(10)\rightarrow
SO(6)\times SO(4)$), or \LRS string model
($SO(10)\rightarrow SU(3)\times SU(2)^2\times U(1)$),
and the \SLM string model ($SO(10)\rightarrow SU(3)\times
SU(2)\times U(1)^2$) or SU421 string model ($SO(10)\rightarrow SU(4)\times
SU(2)\times U(1)$),
the bi--doublet splitting problem
favors the later choices. The \SLM and SU421 string models
provide a stringy solution both to the doublet--triplet splitting
problem, as well as to the bi--doublet splitting problem.
However, as we discuss below the NAHE--based free fermionic models
do not produce viable SU421 string models.

\setcounter{footnote}{0}
\section{SU421 free fermionic models}\label{su421ffm}

A model in the free fermionic formulation \cite{fff} is constructed by
choosing a consistent set of boundary condition basis vectors.
The basis vectors, $\mb_k$, span a finite
additive group $\Xi=\sum_k{{n_k}{\mb_k}}$
where $n_k=0,\cdots,{{N_{z_k}}-1}$.
The physical massless states in the Hilbert space of a given sector
$\malpha\in{\Xi}$, are obtained by acting on the vacuum with
bosonic and fermionic operators and by
applying the generalized GSO projections. The $U(1)$
charges, $Q(f)$, for the unbroken Cartan generators of the four
dimensional gauge group are in one
to one correspondence with the $U(1)$
currents ${f^*}f$ for each complex fermion f, and are given by:
\begin{equation}
{Q(f) = {1\over 2}\alpha(f) + F(f)},
\label{u1charges}
\end{equation}
where $\alpha(f)$ is the boundary condition of the world--sheet fermion $f$
in the sector $\malpha$, and
$F_\alpha(f)$ is a fermion number operator counting each mode of
$f$ once (and if $f$ is complex, $f^*$ minus once).
For periodic fermions,
$\alpha(f)=1$, the vacuum is a spinor representation of the Clifford
algebra of the corresponding zero modes.
For each periodic complex fermion $f$
there are two degenerate vacua ${\vert +\rangle},{\vert -\rangle}$ ,
annihilated by the zero modes $f_0$ and
${{f_0}^*}$ and with fermion numbers  $F(f)=0,-1$, respectively.

The realistic models in the free fermionic formulation are generated by
a basis of boundary condition vectors for all world--sheet fermions
\cite{fsu5,fny,alr,slm,custodial}. The basis is constructed in
two stages. The first stage consists of the NAHE set
\cite{fsu5,nahe,slm}\footnote{The NAHE--set was first
constructed by Nanopoulos, Antoniadis, Hagelin and Ellis  (NAHE)
as a subset of the flipped $SU(5)$ string model\cite{fsu5}. 
Its properties and importance for the phenomenological viability 
of the free fermionic models were discussed in reference
\cite{nahe,slm}.  {\it nahe}=pretty, in Hebrew.},
which is a set of five boundary condition basis
vectors, $\{{{\bf 1},\mS,\mb_1,\mb_2,\mb_3}\}$.
The gauge group after the NAHE set
is $SO(10)\times SO(6)^3\times E_8$ with $N=1$ space--time supersymmetry.
The vector $\mS$ is the supersymmetry generator and the superpartners of
the states from a given sector $\malpha$ are obtained from the sector
$\mS+\malpha$. The space--time vector bosons that generate the gauge group
arise from the Neveu--Schwarz (NS) sector and from the sector $\mzeta
\equiv {\mathbf 1}+\mb_1+\mb_2+\mb_3$.
The NS sector produces the generators of
$SO(10)\times SO(6)^3\times SO(16)$. The sector
$\mzeta$
produces the spinorial $\bf 128$ of $SO(16)$ and completes the hidden
gauge group to $E_8$. The vectors $\mb_1$, $\mb_2$ and $\mb_3$
produce 48 spinorial $\bf 16$'s of $SO(10)$, sixteen from each sector $\mb_1$,
$\mb_2$ and $\mb_3$. The vacuum of these sectors contains eight periodic
worldsheet fermions, five of which produce the charges under the
$SO(10)$ group, while the remaining three periodic fermions
generate charges with respect to the flavor symmetries. Each of the
sectors $\mb_1$, $\mb_2$ and $\mb_3$ is charged with respect to a
different set of flavor quantum numbers, $SO(6)_{1,2,3}$.

The NAHE set divides the 44 right--moving and 20 left--moving real internal
fermions in the following way: ${\bar\psi}^{1,\cdots,5}$ are complex and
produce the observable $SO(10)$ symmetry; ${\bar\phi}^{1,\cdots,8}$ are
complex and produce the hidden $E_8$ gauge group;
$\{{\bar\eta}^1,{\bar y}^{3,\cdots,6}\}$, $\{{\bar\eta}^2,{\bar y}^{1,2}
,{\bar\omega}^{5,6}\}$, $\{{\bar\eta}^3,{\bar\omega}^{1,\cdots,4}\}$
give rise to the three horizontal $SO(6)$ symmetries. The left--moving
$\{y,\omega\}$ states are also divided into the sets $\{{y}^{3,\cdots,6}\}$,
$\{{y}^{1,2}
,{\omega}^{5,6}\}$, $\{{\omega}^{1,\cdots,4}\}$. The left--moving
$\chi^{12},\chi^{34},\chi^{56}$ states carry the supersymmetry charges.
Each sector $\mb_1$, $\mb_2$ and $\mb_3$ carries periodic boundary conditions
under $(\psi^\mu\vert{\bar\psi}^{1,\cdots,5})$ and one of the three groups:
$(\chi_{12},\{y^{3,\cdots,6}\vert{\bar y}^{3,\cdots6}\},{\bar\eta}^1)$,
$(\chi_{34},\{y^{1,2},\omega^{5,6}\vert{\bar y}^{1,2}{\bar\omega}^{5,6}\},
{\bar\eta}^2)$,
$(\chi_{56},\{\omega^{1,\cdots,4}\vert{\bar\omega}^{1,
\cdots4}\},{\bar\eta}^3)$.

The second stage of the basis construction consist of adding three
additional basis vectors to the NAHE set.
Three additional vectors are needed to reduce the number of generations
to three, one from each sector $\mb_1$, $\mb_2$ and $\mb_3$.
One specific example is given in Table (\ref{model1}).
The choice of boundary
conditions to the set of real internal fermions
${\{y,\omega\vert{\bar y},{\bar\omega}\}^{1,\cdots,6}}$
determines the low energy properties, such as the number of generations,
Higgs doublet--triplet splitting and Yukawa couplings.

The $SO(10)$ gauge group is broken to one of its subgroups
$SU(5)\times U(1)$, $SO(6)\times SO(4)$ or
$SU(3)\times SU(2)\times U(1)^2$ by the assignment of
boundary conditions to the set ${\bar\psi}^{1\cdots5}_{1\over2}$:

\beqn
&1.&b\{{{\bar\psi}_{1\over2}^{1\cdots5}}\}=
\{{1\over2}{1\over2}{1\over2}{1\over2}
{1\over2}\}\Rightarrow SU(5)\times U(1),\label{su51breakingbc}\\
&2.&b\{{{\bar\psi}_{1\over2}^{1\cdots5}}\}=\{1 1 1 0 0\}
  \Rightarrow SO(6)\times SO(4).
\label{su51so64breakingbc}
\eeqn

To break the $SO(10)$ symmetry to
$SU(3)_C\times SU(2)_L\times
U(1)_C\times U(1)_L$\footnote{$U(1)_C={3\over2}U(1)_{B-L};
U(1)_L=2U(1)_{T_{3_R}}.$}
both steps, 1 and 2, are used, in two separate basis vectors.
The breaking pattern
$SO(10)\rightarrow SU(3)_C\times SU(2)_L\times SU(2)_R \times U(1)_{B-L}$
is achieved by the following assignment in two separate basis
vectors
\beqn
&1.&b\{{{\bar\psi}_{1\over2}^{1\cdots5}}\}=\{1 1 1 0 0\}
  \Rightarrow SO(6)\times SO(4),\\
&2.&b\{{{\bar\psi}_{1\over2}^{1\cdots5}}\}=
\{{1\over2}{1\over2}{1\over2}00\}\Rightarrow SU(3)_C\times U(1)_C
\times SU(2)_L\times SU(2)_R.
\label{su3122breakingbc}
\eeqn

Similarly, the breaking pattern
$SO(10)\rightarrow SU(4)_C\times SU(2)_L\times U(1)_R$
is achieved by the following assignment in two separate basis
vectors

\beqn
&1.& b\{{{\bar\psi}_{1\over2}^{1\cdots5}}\}=\{1 1 1 0 0\}
  \Rightarrow SO(6)\times SO(4),\\
&2.& b\{{{\bar\psi}_{1\over2}^{1\cdots5}}\}=
\{000{1\over2}{1\over2}\}\Rightarrow SU(4)_C\times
SU(2)_L\times U(1)_R.
\label{su421breakingbc}
\eeqn

We comment here that
a recurring feature of some of the three generation free fermionic heterotic
string models is the emergence of a combination of the basis vectors
which extend the NAHE set,
\begin{equation}
\mX=n_\malpha\malpha+n_\mbeta\mbeta+n_\mgamma\mgamma\label{xcomb},
\end{equation}
for which $\mX_L\cdot \mX_L=0$ and $\mX_R\cdot \mX_R\ne0$. Such a
combination may produce additional space--time vector
bosons, depending on the choice of GSO phases.
These additional space--time vector bosons
enhance the four dimensional gauge group.
This situation is similar to the presence of the combination
of the NAHE set basis vectors ${\mathbf 1}+\mb_1+\mb_2+\mb_3$, which
enhances the hidden gauge group, at the level of the NAHE set,
{}from $SO(16)$ to $E_8$.
In the free fermionic models this type of
gauge symmetry enhancement in the observable sector is,
in general, family universal and is intimately related to the
$\IZ_2\times \IZ_2$ orbifold structure which underlies the
realistic free fermionic models. Such
enhanced symmetries were shown to forbid proton decay
mediating operators to all orders of nonrenormalizable
terms \cite{custodial}. Below we discuss examples of
models with and without gauge enhancement.

The SU421 symmetry breaking pattern induced by the
boundary condition assignment given in eq. (\ref{su421breakingbc})
has an important distinction from the previous symmetry breaking patterns.
As in the previous cases, since it involves a breaking of an
$SO(2n)$ group to $SU(n)\times U(1)$ it contains 1/2 boundary
conditions. As discussed above the observable and hidden
non--Abelian gauge groups arise from the sets of complex world--sheet
fermions $\{{\bar\psi}^{1,\cdots,5}{\bar\eta}^{1,\cdots,3}\}$ and
$\{{\bar\phi}^{1,\cdots,8}\}$, respectively. The breaking pattern
(\ref{su421breakingbc}) entails an assignment of $1/2$ boundary
condition to two complex fermions in the observable set,
whereas the symmetry breaking patterns in eqs.
(\ref{su51breakingbc},\ref{su3122breakingbc})
involve three
such assignments. On the other hand, the modular invariance
rules \cite{fff} for the product $\mb_j\cdot\mgamma$, where $\mb_j$ are
the NAHE set basis vectors and $\mgamma$ is the basis vector that
contains the $1/2$ boundary conditions, enforces that no other
complex fermion from the observable set has $1/2$ boundary conditions.
Additionally, the constraint on the product $\mgamma\cdot\mgamma$ imposes
that either 8 or 12 complex fermions have $1/2$ boundary conditions.
Since, as we saw, only two can have such boundary conditions
from the observable set, it implies that six and only six from
the hidden set must have $1/2$ boundary conditions. This is
in contrast to the other cases that allow assignment of
twelve $1/2$ boundary conditions in the basis vector $\mgamma$.
The consequence of having only eight $1/2$ boundary conditions in the
basis vector $\mgamma$ is the appearance of additional sectors
that may lead to enhancement of the four dimensional gauge group.
Below we discuss several other important distinctions of
this symmetry breaking pattern as compared to the
previous cases.

\section{SU421 model without enhanced symmetry}\label{noes}

As our first example of a SU421
free fermionic heterotic string model
we consider Model 1, specified below.
The boundary conditions of the three basis vectors
which extend the NAHE set are shown in Table (\ref{model1}).
Also given in Table (\ref{model1}) are the pairings of
left-- and right--moving real fermions from the set
$\{y,\omega|{\bar y},{\bar\omega}\}$. These fermions are
paired to form either complex, left-- or right--moving fermions,
or Ising model operators, which combine a real left--moving fermion with
a real right--moving fermion. The generalized GSO coefficients
determining the physical massless states of Model 1 appear in
matrix (\ref{phasesmodel1}).
\vskip 0.4truecm

\LRS Model 1 Boundary Conditions:
\beqn
 &\begin{tabular}{c|c|ccc|c|ccc|c}
 ~ & $\psi^\mu$ & $\chi^{12}$ & $\chi^{34}$ & $\chi^{56}$ &
        $\bar{\psi}^{1,...,5} $ &
        $\bar{\eta}^1 $&
        $\bar{\eta}^2 $&
        $\bar{\eta}^3 $&
        $\bar{\phi}^{1,...,8} $ \\
\hline
\hline
  ${\malpha}$  &  0 & 0&0&0 & 1~1~1~0~0 & 0 & 0 & 0 &1~1~1~1~0~0~0~0 \\
  ${\mbeta}$   &  0 & 0&0&0 & 1~1~1~0~0 & 0 & 0 & 0 &1~1~1~1~0~0~0~0 \\
  ${\mgamma}$  &  0 & 0&0&0 & 0~0~0~${1\over2}$~${1\over2}$& 0 & 0 & 0 &
  ${1\over2}$~${1\over2}$~${1\over2}$~${1\over2}$~
  ${1\over2}$~${1\over2}$~0~0
\end{tabular}
   \nonumber\\
   ~  &  ~ \nonumber\\
   ~  &  ~ \nonumber\\
     &\begin{tabular}{c|c|c|c}
 ~&   $y^3{y}^6$
      $y^4{\bar y}^4$
      $y^5{\bar y}^5$
      ${\bar y}^3{\bar y}^6$
  &   $y^1{\omega}^5$
      $y^2{\bar y}^2$
      $\omega^6{\bar\omega}^6$
      ${\bar y}^1{\bar\omega}^5$
  &   $\omega^2{\omega}^4$
      $\omega^1{\bar\omega}^1$
      $\omega^3{\bar\omega}^3$
      ${\bar\omega}^2{\bar\omega}^4$ \\
\hline
\hline
$\malpha$& 1 ~~~ 1 ~~~ 1 ~~~ 0  & 1 ~~~ 1 ~~~ 1 ~~~ 0  & 1 ~~~ 1 ~~~ 1 ~~~ 0 \\
$\mbeta$ & 0 ~~~ 1 ~~~ 0 ~~~ 1  & 0 ~~~ 1 ~~~ 0 ~~~ 1  & 1 ~~~ 0 ~~~ 0 ~~~ 0 \\
$\mgamma$& 0 ~~~ 0 ~~~ 1 ~~~ 1  & 1 ~~~ 0 ~~~ 0 ~~~ 0  & 0 ~~~ 1 ~~~ 0 ~~~ 0 \\
\end{tabular}
\label{model1}
\eeqn

\LRS Model 1 Generalized GSO Coefficients:
\begin{equation}
{\bordermatrix{
 &{\bf 1}&\mS & & {\mb_1}&{\mb_2}&{\mb_3}& & {\malpha}&{\mbeta}&{\mgamma}\cr
{\bf 1}&~~1&~~1 & & -1   &  -1 & -1   & &  ~~1     & ~~1   & ~~1   \cr
           \mS&~~1&~~1 & &~~1   & ~~1 &~~1   & &   -1     &  -1   &  -1   \cr
	      &   &    & &      &     &      & &         &       &       \cr
       {\mb_1}& -1& -1 & & -1   &  -1 & -1   & &   -1     &  -1   & ~~1   \cr
       {\mb_2}& -1& -1 & & -1   &  -1 & -1   & &  ~~1     & ~~1   & ~~1   \cr
       {\mb_3}& -1& -1 & & -1   &  -1 & -1   & &   -1     & ~~1   & ~~i   \cr
	      &   &    & &      &     &      & &          &       &       \cr
     {\malpha}&~~1& -1 & &~~1   &  -1 &~~1   & &  ~~1     & ~~1   & ~~i   \cr
     { \mbeta}&~~1& -1 & & -1   & ~~1 & -1   & &   -1     &  -1   & ~~i   \cr
     {\mgamma}&~~1& -1 & & -1   & ~~1 & -1   & &   -1     & ~~1   & ~~i   \cr}}
\label{phasesmodel1}
\end{equation}

In matrix (\ref{phasesmodel1}) only the entries above the diagonal are
independent and those below and on the diagonal are fixed by
the modular invariance constraints.
Blank lines are inserted to emphasize the division of the free
phases between the different sectors of the realistic
free fermionic models. Thus, the first two lines involve
only the GSO phases of $c{{\{{\bf 1},\mS\}}\choose \ma_i}$. The set
$\{{\bf 1},\mS\}$ generates the $N=4$ model with $\mS$ being the
space--time supersymmetry generator and therefore the phases
$c{\mS\choose{\ma_i}}$ are those that control the space--time supersymmetry
in the superstring models. Similarly, in the free fermionic
models, sectors with periodic and anti--periodic boundary conditions,
of the form of $\mb_i$, produce the chiral generations.
The phases $c{\mb_i\choose \mb_j}$ determine the chirality
of the states from these sectors.

In the free fermionic models
the basis vectors $\mb_i$ are those that respect the $SO(10)$ symmetry
while the vectors denoted by Greek letters are those that break the
$SO(10)$ symmetry.
As the Standard Model matter states arise from sectors which
preserve the $SO(10)$ symmetry, the phases that fix
the Standard Model charges are, in general,
the phases $c{\mb_i\choose{\ma_i}}$. On the other hand,
the basis vectors of the form $\{\malpha,\mbeta,\mgamma\}$ break the
$SO(10)$ symmetry. The phases associated with these basis vectors
are associated with exotic physics, beyond the Standard Model.
These phases, therefore, also affect the final four dimensional
gauge symmetry.

The final gauge group in Model 1 arises
as follows: In the observable sector the NS boundary conditions
produce gauge group generators for
\beq
SU(4)_C\times SU(2)_L\times U(1)_R\times U(1)_{1,2}\times SU(2)_3\times
U(1)_{4,5}\times SU(2)_6.
\eeq
Here the flavor $SU(2)_{3,6}$ symmetries are generated by
$\{{\bar\eta}^3{\bar\zeta}^3\}$ where ${\bar\zeta}^3=1/\sqrt{2}
({\bar w}^2+i{\bar w}^4)$. In previous free fermionic models
this group factor breaks to $U(1)^2$, but this is an artifact
of the specific model considered in eq. (\ref{model1}), and
is not a generic feature of SU421 models.
Thus, the $SO(10)$ symmetry is broken to
$SU(4)\times SU(2)_L\times U(1)_R$, as discussed above,
where,
\begin{equation}
U(1)_R={\rm Tr}\, U(2)_L~\Rightarrow~Q_R=
			 \sum_{i=4}^5Q({\bar\psi}^i).
\label{u1r}
\end{equation}
The flavor $SO(6)^3$ symmetries are broken to $U(1)_{1,2}\times SU(2)_3
\times U(1)_{4,5}\times SU(2)_6$.
In the hidden sector the NS boundary conditions produce the generators of
\beq
SU(4)_{H}\times SU(2)_{H1}\times SU(2)_{H2}\times SU(2)_{H3}\times
U(1)_{7,8}
\label{nshiden}
\eeq
where $SU(2)_{H1,H2}$ and $SU(2)_{H3}$ arise from the complex
world--sheet fermions
$\{{\bar\phi}^7{\bar\phi}^8\}$ and $\{{\bar\phi}^5{\bar\phi}^6\}$,
respectively; and $U(1)_{7}$ and $U(1)_{8}$
correspond to the combinations of world--sheet charges
\begin{eqnarray}
Q_{7}&=&\sum_{i=5}^6Q({\bar\phi})^i.\label{qh2model1}\\
Q_{8}&=&\sum_{i=1}^4Q({\bar\phi})^i,\label{qh1model1}
\end{eqnarray}

As we discussed in section (\ref{su421ffm}) the SU421 models
contain additional sectors that may produce space--time
vector bosons and enhance the four dimensional gauge group.
In the model of eq. (\ref{model1}) these are the sectors
$2\mgamma$, $\mzeta_1\equiv1+\mb_1+\mb_2+\mb_3$ and
 $\mzeta_2\equiv1+\mb_1+\mb_2+\mb_3+2\mgamma$. However, due to the
choice of one--loop phases in eq. (\ref{phasesmodel1})
all the additional vector bosons from these sectors are
projected out by the GSO projections and there is therefore
no gauge enhancement from these sectors in this model.
These sectors are generic in the SU421 models.

In addition to the graviton, dilaton, antisymmetric sector and
spin--1 gauge bosons, the NS sector gives two pairs
of color triplets, transforming as (6,1,0) under
$SU(4)_C\times SU(2)_L\times U(1)_R$;
three quadruplets of $SO(10)$ singlets with
$U(1)_{1,2,3}$ charges; and three singlets of the entire four
dimensional gauge group.
The states from the sectors $\mb_j$ $(j=1,2,3)$ produce
the three light twisted generations. These states and their
decomposition under the entire gauge group are shown in
Appendix A. The remaining massless states and their quantum numbers
also appear in Appendix A .

A distinct feature of the SU421 NAHE--based free fermionic models
is with respect to the states from the twisted sectors $\mb_{1,2,3}$.
As discussed in section (\ref{su421ffm}), in the
NAHE--based free fermionic models the twisted sectors $\mb_j$
produce the three chiral 16 of $SO(10)$ decomposed under the
final $SO(10)$ subgroup. In the case of the FSU5, PS and SLM models
the 2$\mgamma$ projection fixes the sign of the charge under $U(1)_{1,2,3}$
to be either plus or minus. The consequence is that in these models there
exists a combination of the flavor symmetries, which is anomalous.
Additionally, in these models the 2$\mgamma$ projection does not break the
$SO(10)$ symmetry. By contrast, in the left--right symmetric
models of ref. \cite{cfs}, and in the SU421 models, the
2$\mgamma$ projection breaks the $SO(10)$ symmetry.
In the FSU5, PS, SLM and LRS models, however, the 2$\mgamma$
projection selects complete 16 representations of $SO(10)$ that
carry different charges under the flavor $U(1)$ symmetries,
{\it i.e.} in the orbifold language they may attach to different
fixed points. By contrast, however, in the case of the SU421 models
the 2$\mgamma$ projection selects partial fillings of the 16 representation
of $SO(10)$, {\it i.e.} either the left or right--handed fields.
The consequence is that the SU421 models do not produce three
complete Standard Model generations and hence the models
are not realistic.

As discussed above the existence of an anomalous $U(1)$ symmetry is a
general outcome of free fermionic models. However, in the case
of LRS free fermionic models there exist three chiral generation
models that are completely free of Abelian and non--Abelian anomalies.
This is again a consequence of the 2$\mgamma$ projection.
In the case of the FSU5, PS and SLM models the 2$\mgamma$ vector
breaks the $E_8\times E_8$ gauge group to $SO(16)\times SO(16)$.
The result of the $Z_2\times Z_2$ orbifold projection
is then to break the observable $SO(16)$ gauge symmetry
to $SO(10)\times U(1)^3$. Alternatively, we can view the $Z_2\times
Z_2$ projection as breaking $E_8\rightarrow E_6\times U(1)^2$,
and the 2$\mgamma$ projection induces the breaking $E_6\rightarrow
SO(10)\times U(1)_A$, where $U(1)_A$ is the anomalous $U(1)$
combination.
%
However,
the \LRS free fermionic string models
do not start with the $N=4$ $E_8\times E_8$ or $SO(16)\times SO(16)$ vacua.
Rather, in this case the starting $N=4$ vacua
has $SO(16)\times E_7\times E_7$ gauge group.
The $Z_2\times Z_2$ orbifold acts as
in the previous models.
%
The appearance of an anomalous $U(1)$ symmetry in the FSU5, PS and
SLM string models is therefore tied to the breaking of $E_6\rightarrow
SO(10)\times U(1)_A$, and arises from the chiral spectrum. In the
LRS string models the charges of the chiral spectrum under
$U(1)_A$ cancels sector by sector and therefore the $U(1)_{1,2,3}$
are anomaly free and are absent from the anomalous $U(1)$ combination.
On the other hand the additional $U(1)$ symmetries that arise from
the Narain lattice can be anomalous in the LRS models and
indeed LRS string models that do contain an anomalous $U(1)$ were
presented in ref. \cite{cfs,ccf}.
As we demonstrate below the SU421 models give rise to anomaly free
models as well as to models that contain an anomalous $U(1)$ symmetry.
However, in contrast to the LRS symmetric models the cancellation of the
anomalous $U(1)$ charges of the chiral spectrum
may occur between different sectors rather than sector by sector.

\section{Models with enhanced non--Abelian symmetries}\label{es}

We next turn to our second example, Model 2. The boundary condition basis
vectors and one--loop phases, which define the model, are given in
Table~(\ref{model2}) and matrix (\ref{phasesmodel2}), respectively.
\vskip 0.4truecm

\LRS Model 2 Boundary Conditions:
\beqn
 &\begin{tabular}{c|c|ccc|c|ccc|c}
 ~ & $\psi^\mu$ & $\chi^{12}$ & $\chi^{34}$ & $\chi^{56}$ &
        $\bar{\psi}^{1,...,5} $ &
        $\bar{\eta}^1 $&
        $\bar{\eta}^2 $&
        $\bar{\eta}^3 $&
        $\bar{\phi}^{1,...,8} $ \\
\hline
\hline
 ${\malpha}$  &  0 & 0&0&0 & 1~1~1~0~0 & 0 & 0 & 0 &1~1~1~1~0~0~0~0 \\
 ${\mbeta}$   &  0 & 0&0&0 & 1~1~1~0~0 & 0 & 0 & 0 &1~1~1~1~0~0~0~0 \\
 ${\mgamma}$  &  0 & 0&0&0 & 0~0~0~${1\over2}$~${1\over2}$ & 0 & 0 & 0 &
	${1\over2}$  ${1\over2}$  ${1\over2}$  ${1\over2}$
	${1\over2}$  ${1\over2}$  0~0
\end{tabular}
   \nonumber\\
   ~  &  ~ \nonumber\\
   ~  &  ~ \nonumber\\
     &\begin{tabular}{c|c|c|c}
 ~&   $y^3{y}^6$
      $y^4{\bar y}^4$
      $y^5{\bar y}^5$
      ${\bar y}^3{\bar y}^6$
  &   $y^1{\omega}^5$
      $y^2{\bar y}^2$
      $\omega^6{\bar\omega}^6$
      ${\bar y}^1{\bar\omega}^5$
  &   $\omega^2{\omega}^4$
      $\omega^1{\bar\omega}^1$
      $\omega^3{\bar\omega}^3$
      ${\bar\omega}^2{\bar\omega}^4$ \\
\hline
\hline
$\malpha$& 0 ~~~ 0 ~~~ 1 ~~~ 1 & 1 ~~~ 0 ~~~ 0 ~~~ 0  & 0 ~~~ 1 ~~~ 0 ~~~ 1 \\
$\mbeta$ & 1 ~~~ 0 ~~~ 0 ~~~ 0 & 0 ~~~ 0 ~~~ 1 ~~~ 1  & 0 ~~~ 0 ~~~ 1 ~~~ 1 \\
$\mgamma$& 0 ~~~ 1 ~~~ 0 ~~~ 1 & 0 ~~~ 1 ~~~ 0 ~~~ 1  & 1 ~~~ 0 ~~~ 0 ~~~ 1 \\
\end{tabular}
\label{model2}
\eeqn

\LRS Model 2 Generalized GSO Coefficients:
\begin{equation}
{\bordermatrix{
              &{\bf 1}&\mS & & {\mb_1}&{\mb_2}&{\mb_3}&
&{\malpha}&{\mbeta}&{\mgamma}\cr
       {\bf 1}&~~1&~~1 & & -1   &  -1 & -1  & & ~~1     & ~~1   & ~~1   \cr
           \mS&~~1&~~1 & &~~1   & ~~1 &~~1  & &  -1     &  -1   &  -1   \cr
	      &   &    & &      &     &     & &         &       &       \cr
       {\mb_1}& -1& -1 & & -1   &  -1 & -1  & & ~~1     &  -1   &  -1   \cr
       {\mb_2}& -1& -1 & & -1   &  -1 & -1  & &  -1     &  -1   &  -1   \cr
       {\mb_3}& -1& -1 & & -1   &  -1 & -1  & & ~~1     & ~~1   & ~~i   \cr
	      &   &    & &      &     &     & &         &       &       \cr
     {\malpha}&~~1& -1 & &~~1   & ~~1 &~~1  & &  -1     & ~~1   &  -1   \cr
     {\mbeta} &~~1& -1 & &~~1   &  -1 &~~1  & & ~~1     &  -1   & ~~1   \cr
     {\mgamma}& -1& -1 & &~~1   & ~~1 & -1  & &  -1     & ~~1   &  -1   \cr}}
\label{phasesmodel2}
\end{equation}

The total gauge group of Model 2 arises as follows.
In the observable sector the NS boundary conditions produce the
generators of
$SU(4)_C \times SU(2)_L\times U(1)_R\in SO(10))
\times U(1)_{1,2,3}\times U(1)_{4,5,6}$,
while in the hidden sector
the NS boundary conditions produce the generators of
\beq
SU(4)_{H}\times SU(2)_{H_1}\times SU(2)_{H_2}\times
SU(2)_{H_3}\times U(1)_7\times U(1)_8\, .
\eeq
$U(1)_{7}$ and $U(1)_{8}$
correspond to the combinations of the world--sheet charges
given in eqs. (\ref{qh1model1}) and (\ref{qh2model1}), respectively.

Model 2 contains two combinations
of non--NAHE basis vectors with $\mX_L\cdot \mX_L=0$, which
therefore may give rise to additional space--time vector bosons.
The first is the sector ${2\mgamma}$.
The second arises from the vector combination given by $\mzeta+2\mgamma$,
where $\mzeta\equiv1+\mb_1+\mb_2+\mb_3$.
Both sectors arise from the
NAHE set basis vectors plus $2\mgamma$ and are therefore
independent of the assignment of periodic boundary
conditions in the basis vectors $\malpha$, $\mbeta$ and $\mgamma$.
Both are therefore generic for the
pattern of symmetry breaking $SO(10)\rightarrow
SU(4)_C\times  SU(2)_L\times U(1)_R$,
in NAHE based models.

In Model 1 the additional space--time vector bosons from both
sectors are projected out and therefore there is no gauge
enhancement. In Model 2 all the
space--time vector bosons from the sector $2\mgamma$
are projected out by the GSO projections and therefore
give no gauge enhancement from this sector.
The sector $\mzeta+2\mgamma$ may, or may not, give
rise to additional space--time vector bosons, depending on
the choice of GSO phase
\beq
c{\mgamma\choose \mb_3}=\pm1,
\label{b3gammaphase}
\eeq
where with the $+1$ choice all the additional vector bosons are
projected out,
whereas the $-1$ choice gives rise to additional space--time gauge bosons
which are charged with respect to the
$SU(2)_L\times SU(2)_{H_1}$ groups. This enhances the
$SU(2)_L\times SU(2)_{H_1}$ group to $SO(5)$.
Thus, in this case,
the full massless spectrum transforms under the final gauge group,
$SU(4)_C\times SO(5)\times U(1)_R\times
U(1)_{1,2,3}\times U(1)_{4,5,6}\times
SU(4)_{H}\times SU(2)_{H_2}\times SU(2)_{H_3}\times
U(1)_{7,8}$.

In addition to the graviton,
dilaton, antisymmetric sector and spin--1 gauge bosons,
the NS sector gives rise to
three quadruplets of $SO(10)$ singlets with
$U(1)_{1,2,3}$ charges; and three singlets of the entire four
dimensional gauge group.
The states from the sectors $\mb_j\oplus \mzeta+2\mgamma~(j=1,2,3)$ produce
the three light generations. The states from these sectors and their
decomposition under the entire gauge group are shown in Appendix B.

\section{Anomalous $U(1)$}\label{anomalousu1}

A general property of the realistic free fermionic heterotic string models,
which is also shared by many other superstring vacua, is the
existence of an ``anomalous" $U(1)$.
The presence of an Abelian anomalous symmetry in superstring
derived models yields many desirable phenomenological consequences
{}from the point of view of the effective low energy field theory.
Indeed, the existence of such an anomalous $U(1)$ symmetry
in string derived models has inspired vigorous attempts to
understand numerous issues, relevant for the observable phenomenology,
including: the fermion mass spectrum, supersymmetry breaking
cosmological implications, and more. From the perspective
of string phenomenology an important function of the
anomalous $U(1)$ is to induce breaking and rank reduction
of the four dimensional gauge group. In general, the existence
of an anomalous $U(1)$ in a string model implies that the
string vacuum is unstable and must be shifted
to a stable point in the moduli space.
This arises because, by the Green--Schwarz
anomaly cancellation mechanism, the anomalous $U(1)$ gives rise
to a Fayet--Iliopoulos term which breaks supersymmetry.
Supersymmetry is restored and the vacuum is stabilized by sliding the
vacuum along flat $F$ and $D$ directions. This is achieved by assigning
non--vanishing VEVs to some scalar fields in the massless string spectrum.

An important issue in string phenomenology is therefore to
understand what are the general conditions for the appearance
of an anomalous $U(1)$ and under what conditions an anomalous
$U(1)$ is absent. The previously studied realistic free fermionic
string models that include the \FSUc, \PSc, and \SLM
types, have always contained an anomalous $U(1)$ symmetry.
In contrast, in ref. \cite{cfs} it was shown that the \LRS models
also give rise to vacua in which all the $U(1)$ symmetries in the four
dimensional gauge group are anomaly free, as well as to vacua that
do contain an anomalous $U(1)$ symmetry.
The distinction between the different cases, and the properties
of the models that resulted in the presence, or the absence, of an
anomalous $U(1)$ symmetry were discussed in ref. \cite{cfs}.
Next we examine the presence of an anomalous $U(1)$ symmetry
in the SU421 string models.

For completeness we first discuss the case of the previously studied
free fermionic models, {\it i.e.}
the \FSUc, the \PSc, the \SLM and the \LRS string models.
The question of the anomalous $U(1)$ symmetry in string models,
in general, and in the free fermionic models, in particular,
was studied in some detail
in ref.~\cite{kn,cf1}. The anomalous $U(1)$ in the free fermionic models
is in general a combination of two distinct kinds of world--sheet $U(1)$
currents,
those generated by ${\bar\eta}^j$ and those generated by the additional
complexified fermions from the set $\{{\bar y},{\bar\omega}\}^{1,\cdots,6}$.
The trace of the $U(1)$ charges of the entire massless string spectrum
can then be non--vanishing under some of these world--sheet $U(1)$ currents.
One combination of these $U(1)$ currents then becomes the anomalous
$U(1)$, whereas all the orthogonal combinations are anomaly free.
To understand the origin of the anomalous $U(1)$ in the realistic
free fermionic models, it is instructive to consider the contributions
{}from the two types of world--sheet $U(1)$ currents separately.

In ref.~\cite{cf1} it was shown that the anomalous $U(1)$
in the realistic free fermionic models can be seen to arise
due to the breaking of the world--sheet supersymmetry from
(2,2) to (2,0). Consider the set of boundary condition
basis vectors $\{{\bf1},\mS, \mzeta,\mX,\mb_1,\mb_2\}$ \cite{cf1},
which produces (for an appropriate choice of the GSO phases)
the model with $SO(4)^3\times E_6\times U(1)^2\times E_8$ gauge
group. It was shown that if we choose the GSO phases such
that $E_6\rightarrow SO(10)\times U(1)$, the $U(1)$ in the
decomposition of $E_6$ under $SO(10)\times U(1)$ becomes
the anomalous $U(1)$. This $U(1)$ is produced by the
combination of world--sheet currents ${\bar\eta}^1{\bar\eta}^{1^*}+
{\bar\eta}^2{\bar\eta}^{2^*}+{\bar\eta}^3{\bar\eta}^{3^*}$.
We can view all of the realistic \FSUc, \PSc, and \SLM free fermionic
string models as being related to this $SO(4)^3\times
E_6\times U(1)^2\times E_8$ string vacuum. This combination of
$U(1)$ currents therefore contributes to the anomalous $U(1)$
in all the realistic free fermionic models with
\FSUc, \PSc, or \SLM gauge groups.

The existence of the anomalous $U(1)$ in the \FSUc, \PSc, or \SLMc,
and its absence in the \LRS string models
can be traced to different $N=4$ string vacua in four
dimensions. While in the $E_6$ model one starts with
an $N=4$ $SO(12)\times E_8\times E_8$ string vacua,
produced by the set $\{{\bf1},\mS,\mX,\mzeta\}$ \cite{cf1},
we can view the \FSUc, \PSc, and \SLM string models as
starting from an $N=4$ $SO(12)\times SO(16)\times SO(16)$ string vacua.
In this case the two spinorial representations from the sectors
$\mX$ and $\mzeta$, that complete
the adjoint of $SO(16)\times SO(16)$ to $E_8\times E_8$, are
projected out by the choice of the GSO projection phases.
The subsequent projections, induced by the basis vectors
$\mb_1$ and $\mb_2$, which correspond to the $\IZ_2\times \IZ_2$
orbifold twistings, then operate identically in the two models,
producing in one case the $E_6$, and in the second case
the $SO(10)\times U(1)$, gauge groups, respectively.
The important point, however, is that both cases preserve
the ``standard embedding'' structure which splits the
observable and hidden sectors. The important set in this
respect is the set $\{{\bf1},\mS,\mX,\mzeta\}$, where $\mX$
has periodic boundary conditions for $\{{\bar\psi}^{1,\cdots,5},
{\bar\eta}^1,{\bar\eta}^2,{\bar\eta}^3\}$. The choice of
the phase $c{\mX\choose\mzeta}=\pm1$ fixes the vacuum to
$E_8\times E_8$ or $SO(16)\times SO(16)$.

In contrast,
the \LRS free fermionic string models do not start
with the $N=4$ $E_8\times E_8$ or $SO(16)\times SO(16)$ vacua.
Rather, in this case the starting $N=4$ vacua can be seen to arise
{}from the set of boundary condition basis vectors
$\{{\bf1},\mS,2\mgamma,\mzeta\}$.
Starting with this set and with the choice of GSO projection phases
\begin{equation}
{\bordermatrix{
              &{\bf 1}& \mS & & \mzeta    & 2\mgamma \cr
       {\bf 1}&~~1&~~1 & & -1   &  -1     \cr
           \mS&~~1&~~1 & & -1   &  -1     \cr
	      &   &    & &      &         \cr
       \mzeta & -1& -1 & &~~1   & ~~1     \cr
     2\mgamma & -1& -1 & & -1   & ~~1     \cr}}\, ,
\label{phasesn4model}
\end{equation}
the resulting string vacua has $N=4$ space--time supersymmetry
with $SO(16)\times E_7\times E_7$ gauge group.
The sectors $\mb_1$ and $\mb_2$ are then added as
in the previous models. The \LRS string models therefore
do not preserve the ``standard embedding'' splitting between the
observable and hidden sectors. This is the first basic
difference between the \FSUc, \PSc, or \SLMc,
and the \LRS free fermionic models.

Now turn to the case of the three generation models. The chirality
of the generations from the sectors $\mb_j$ $(j=1,2,3)$ is induced
by the projection which breaks $N=2\rightarrow N=1$ space--time
supersymmetry. Chirality for the generations is therefore fixed by the GSO
projection phase $c{\mb_i\choose{\mb_j}}$ with $i\ne j$.
On the other hand,
generation charges under $U(1)_j$ are fixed
by the $\mX$ projection in the $E_6$ model, by
the projection induced by the vector $2\mgamma$
of the \FSUc, \PSc, and \SLM string models, or by the vector
$2\mgamma$ of the \LRS string models. The difference
is that in the case of the \FSUc, \PSc, and \SLM string models
the $2\mgamma$ projection fixes the same sign for the
$U(1)_j$ charges of the states from the sectors $\mb_j$.
In contrast, in the \LRS free fermionic models the corresponding
$2\mgamma$ projection fixes one sign for the $(Q_R+L_R)_j$
states and the opposite sign for the $(Q_L+L_L)_j$ states.
The consequence is that the total trace vanishes and the
sectors $\mb_j$ do not contribute to the trace of the $U(1)_j$ charges.
This is in fact the reason that \LRS free fermionic
models can appear without an anomalous $U(1)$.

The existence of \LRS free fermionic string models
without an anomalous $U(1)$ does not preclude the possibility of
other \LRS models with an anomalous $U(1)$.
Model 3 of ref. \cite{cfs} contains three anomalous
$U(1)$ symmetries, of which
one combination,
\beq
U(1)_A=U_4+U_5+U_6\label{u1a},
\eeq
is anomalous, while two orthogonal combinations
are anomaly free. In this model the anomalous
$U(1)$'s correspond to $U(1)$ symmetries which
arise from the additional complexified world--sheet fermions
in the set $\{{\bar y},{\bar\omega}\}^{1,\cdots,6}$.
This is in agreement with the observation that the
$U(1)_{j=1,2,3}$, which are generated by the ${\bar\eta}^j$
world--sheet fermions, are anomaly free in the \LRS
free fermionic string models.

In the case of the SU421 models the starting $N=4$ vacua differs
again from the previous cases. In this case the set of boundary condition
basis vectors $\{{\bf1},{\bf S},2\mgamma, \zeta\}$, with $\mgamma$ being that
of eqs. (\ref{model1}, \ref{model2}) is to produce the gauge group
$SO(28)\times E_8$. In this model the NS sector produces the
generators of the adjoint of $SO(24)\times SU(2)\times SU(2)
\times SO(12)\times SU(2)\times SU(2)$; the sectors $\zeta$ and
$2\mgamma$ produce gauge bosons that transform in the $(1,1,2,32,1,1)$
and $(1,1,1,{\overline{32}},2,1)$ representations of the NS gauge
group; and the sector $\zeta+2\mgamma$ produces gauge boson
that transform as $(24,2,1,1,1,2)$ and $(1,1,2,12,2,1)$ hence
completing the gauge group to $SO(28)\times E_8$. The SU421 models
are similar to the \LRS models in the sense that, in contrast to the
case of the \FSU, \PS and \SLM models, the periodic
fermions in the basis vectors $2\mgamma$ and $\zeta$ are not
disjoint. The consequence is that, for either choice
of the GSO phase $c{\zeta\choose{2\mgamma}}$,
the $N=4$ gauge group is the same. Hence, discrete choice
of this phase does not generate an anomalous $U(1)$.
Consequently, both the \LRS $SO(10)$ subgroup
as well as the SU421 one produce models in which all
the $U(1)$ currents are anomaly free. Both cases, however,
can produce models which do contain an anomalous $U(1)$.
However, as discussed above, whereas, in the case of the
\LRS models the cancellation of the $U(1)_j$ $(j=1,2,3)$
is within sectors, in the case of the SU421 models
the cancellation may be between sectors. This difference
is again traced back to the different $N=4$ vacua from
which the models originate, as both cases share the basic
NAHE--set structure. In the model of eq.~(\ref{model1})
we find that the cancellation is within sectors and there
is no anomalous $U(1)$ symmetry. In the model of eq. (\ref{model2})
we find that the situation is more intricate. In this case
the presence or absence of an anomalous $U(1)$ depends on the
GSO phase $c{\mb_3\choose\mgamma}=\pm{i}$ that also fixes the four
dimensional gauge group in this model. We therefore find that the
choice $c{\mb_3\choose\mgamma}=-i$ results in an anomaly free model,
whereas the choice $c{\mb_3\choose\mgamma}=+i$ results in a model
that contains an anomalous $U(1)$ and gauge enhancement from
the sector $\mzeta+2\mgamma$. This situation is reminiscent
of the one in the FSU5, PS and SLM NAHE--based models, in which
the presence or absence of an anomalous $U(1)$ symmetry is
related to the choice of the four dimensional gauge group.
In the anomaly free model we find that the sectors
$\mb_3\oplus \mb_3+\mzeta+2\mgamma$ contribute with opposite sign
to $\tr Q_3=-8+8=0$. Similarly, the sectors $S+\mb_2+\mb_3+\malpha+\mbeta
+\mgamma\oplus1+S+\mb_1+\malpha+\mbeta+\mgamma$ and
$S+\mb_1+\mb_3+\malpha+\mbeta
+\mgamma\oplus1+S+\mb_2+\malpha+\mbeta+\mgamma$ contribute
$\tr Q_3=4-4=0$. With $c{\mb_3\choose\mgamma}=+i$ we find that
these sectors contribute with equal sign to $\tr Q_3.$
In this case the sector $\mb+3\oplus \mb_3+\mzeta+2\mgamma$
together form the complete representations of the
enhanced $SO(5)$ gauge group, and give $\tr Q_3=8+8=16$.
Similarly, the sectors $S+\mb_2+\mb_3+\malpha+\mbeta
+\mgamma\oplus1+S+\mb_1+\malpha+\mbeta+\mgamma$ and
$S+\mb_1+\mb_3+\malpha+\mbeta
+\mgamma\oplus1+S+\mb_2+\malpha+\mbeta+\mgamma$ contribute
$\tr Q_3=12+4=16$. We therefore see that the $U(1)_3$ gauge symmetry
in this model is anomalous with $\tr Q_3=48$.
As is the case in the FSU5, PS and SLM NAHE--based models
the emergence of the anomalous $U(1)$ symmetry is tied to
the choice of four dimensional gauge group and the
GSO phases. We further note that in the anomaly free model
the cancellation is between sectors that belong to the same
orbits and that in the anomalous model, due to the choice
of GSO phases and breaking of the gauge group, becomes anomalous.

An important issue in the string models that contain an anomalous
$U(1)$ symmetry is the existence of supersymmetric flat directions.
Since the anomalous $U(1)$ symmetry generates a Fayet--Iliopoulos
term at one--loop in string perturbation theory \cite{dsw}
supersymmetry is broken, and is restored by assigning non--vanishing
VEVs to massless scalars in the string spectrum along
$F$-- and $D$--flat directions. A vital question is therefore
whether such flat directions exist in a given model.
The flat directions can be classified generically as Abelian,
{\it i.e.} those that use solely scalars that are singlets
of all non--Abelian group factors and non-Abelian flat directions
{\it i.e.} those that also utilize non--Abelian fields.
In the case of the FSU5, PS and SLM NAHE--based models
Abelian flat directions were always found to exist.
In the case of the LRS models on the other hand
it was shown in ref. \cite{ccf} that Abelian flat directions
did not exist in model 3 of ref. \cite{cfs} and only
non--Abelian flat directions exist in the model.

In the case of the SU421 non-anomalous model that we study here,
Abelian flat directions exist, but are rather trivial. Besides the
three uncharged moduli, the only other singlet scalars are two independent
vector-like pairs from the NS sector, $(\phi_1,{\bar\phi}_1)$ with
$Q_1= Q_2=\pm 1$ and $(\phi_2,{\bar\phi}_2)$ with
$Q_1= -Q_2=\pm 1$.  Thus, the elements of the basis set of flat directions
are simply $<\phi_1> = <{\bar\phi}_1>$ and $<\phi_2> = <{\bar\phi}_2>$.

In contrast, in the case of the anomalous SU421 model,
the set of Abelian D-flat directions is more complex.
In addition to the uncharged moduli, there are 32 non-Abelian singlet
scalars that form 16 vector-like pairs, as shown in Appendix B. These 16
vector-like pairs generate an eight element basis set for $D$-flat
directions, as shown in Appendix C.
Four basis elements carry -2 anomalous charge (denoted $Q_3$
in Appendix B) and no charge
under the remaining eight non-anomalous Abelian symmetries. Corresponding
directions with +2 anomalous charge can be formed from these basis
directions by vector-partner
field
substitution. The remaining four basis directions carry neither anomalous
nor non-anomalous Abelian charges. As Appendix B shows, all of the non-Abelian
fields in this
model carry anomalous charge $Q_3 = 0$ or +1/2. Thus, one or more of the the
four basis directions
carrying -2 anomalous charge must appear in the non-perturbatively chosen
$D$-flat direction (for cancellation of the FI $D$-term).

Twelve of the 32 singlet fields whose combinations of VEVs yield the $D$-flat
basis
elements originate in the NS sector. Four of these twelve NS fields
carry -1 anomalous charge while their four corresponding vector partners carry
+1 anomalous
charge. The remaining two pairs of NS vector-like pairs carry no anomalous
charge. The ten additional vector-like pairs of singlets also lack anomalous
charge. Two of these ten pairs originate in the ${\bf S} + 2\gamma$ sector and
the remaining eight pairs
originate in the ${\bf b}_2 + \alpha + 2\gamma$ sector. The net
contributions from the ${\bf S} + 2\gamma$ sector field VEVs is zero for
each basis element.

\oddsidemargin  10.5pt \evensidemargin  10.5pt
\textheight  612pt \textwidth  432pt
\headheight  12pt \headsep  20pt
\footheight  12pt \footskip  40pt

\section{$SO(10)$ breaking patterns}

In this section we present a general argument why free fermionic models 
with $SU(4)\times SU(2)\times U(1)$ cannot in fact be constructed.
For this purpose let us recall that the weight lattice of the 
spinorial $SO(10)$ representation is made of an even number of $\vert-\rangle$
Ramond vacua, out of the total five that make up the $SO(10)$ lattice. 
The sixteen available states can be represented in combinatorial
form
\beq
\left[\left(\matrix{{5}\cr{0}\cr}\right)+\left(\matrix{{5}\cr{2}\cr}\right)+
\left(\matrix{{5}\cr{4}\cr}\right)\right]
\label{spinorial}
\eeq
where the combinatorial factor counts the number of $|-\rangle$
states in the vacuum. Under the breaking pattern $SO(10)\rightarrow
SU(4)\times SU(2)_L\times SU(2)_R$ these decompose as 

\beqn
Q_R & \equiv &
\left[\left(\matrix{{3}\cr{0}\cr}\right)+
      \left(\matrix{{3}\cr{2}\cr}\right)\right]
\left[\left(\matrix{{2}\cr{0}\cr}\right)+
      \left(\matrix{{2}\cr{2}\cr}\right)\right]~=~
({\bar4},1,2) ~=~({\bar4},1,1)_{+1}+({\bar4},1,1)_{-1}~~~~
\label{spinodecoright}\\
Q_L & \equiv &
\left[\left(\matrix{{3}\cr{1}\cr}\right)+
      \left(\matrix{{3}\cr{3}\cr}\right)\right]
\left[\left(\matrix{{2}\cr{1}\cr}\right)\right]~~~~~~~~~~~=~
(4,2,1)
\label{spinorialdecom}
\eeqn
where the last step in (\ref{spinodecoright}) is the decomposition 
under $SU(4)\times SU(2)_L\times U(1)_L$. We recall that the GSO
projection condition on states from a given sector $\alpha\in\Xi$
has the general form \cite{fff}
\begin{equation}
\left\{e^{i\pi({b_i}F_\alpha)}-
{\delta_\alpha}c^*\left(\matrix{\alpha\cr
                 b_i\cr}\right)\right\}\vert{s}\rangle=0
\label{gsoprojections}
\end{equation}
with
\begin{equation}
(b_i{F_\alpha})\equiv\{\sum_{real+complex\atop{left}}-
\sum_{real+complex\atop{right}}\}(b_i(f)F_\alpha(f)),
\label{lorentzproduct}
\end{equation}
where $F_\alpha(f)$ is a fermion number operator counting each mode of
$f$ once (and if $f$ is complex, $f^*$ minus once). For periodic
complex fermions
the fermion number of the
two degenerate vacua $\vert{+}\rangle$, $\vert{-}\rangle$ is
$F(f)=0,-1$ respectively. In Eq. (\ref{gsoprojections}),
$\delta_\alpha=-1$ if $\psi^\mu$ is periodic in the sector $\alpha$,
and $\delta_\alpha=+1$ if $\psi^\mu$ is antiperiodic in the sector $\alpha$.

As can be seen from eq. (\ref{gsoprojections}),
the assignment of rational boundary conditions in the basis vector
that breaks $SU(4)\times SO(4)$ to $SU(4)\times SU(2)\times U(1)$
results in either $Q_L$ or $Q_R$ being selected by the GSO projections.
In the free fermionic models that utilize the $Z_4$ Wilson line breaking,
as is done in the models discussed here, the modular invariance 
constraint on $\gamma\cdot b_j$ imposes that
$\gamma\{{\bar\eta}^1;{\bar\eta}^2;{\bar\eta}^3\}\ne1/2.$
Therefore, unlike the situation in the
$SU(3)\times U(1)\times SU(2)^2$ models 
of refs. \cite{cfs,ccf}, the result is that either
the $Q_L$ or the $Q_R$ states from a given sector $b_j$
are left invariant by the GSO projections. The basic difference
between the $SU(4)\times SU(2) \times U(1)$ model and the
$SU(5)\times U(1)$, $SU(3)\times SU(2)\times U(1)^2$ and
$SU(3)\times U(1)\times SU(2)^2$ models is that the former
uses an even number of rational 1/2 boundary conditions in
$\gamma$ whereas the later uses an odd number. For this reason
the former produces pairs of $Q_L$ or $Q_R$ from the sectors $b_j$
whereas the later produces complete $SO(10)$ multiplets.
Now, in order to produce a phenomenologically viable model
we need three copies of $(Q_L+Q_R)$. However, it is clear that the
SU421 models cannot produce that as, by the argument above, they always 
produce an even number of this combination. One could contemplate
the possibility that the Standard Model representations in such
models arise from non--spinorial sectors. However, this will imply that
the weak hypercharge is not embedded in $SO(10)$ and therefore
$SU(4)\times SU(2)\times U(1)$ will not arise as an $SO(10)$ subgroup.
One can contemplate the possibility of using rational boundary conditions
other than 1/2. However, since the problem here arises because 
of the mismatch of the integral boundary assignment to the set
$\{{\bar\psi}^{1,2,3}\}$ versus the rational boundary condition for the
remaining pair $\{{\bar\psi}^{4,5}\}$, assigning other rational boundary
conditions does not help. Finally, since our argument is solely
based on the weight lattice of the spinorial of $SO(10)$ and
the GSO projection operator, it is completely general and we
expect it to hold in any string construction. Our conclusion 
is that string models with the symmetry breaking pattern 
$SO(10)~\rightarrow~SU(4)\times SU(2)\times U(1)$ are not viable.

\begin{figure}[t]
\centerline{\epsfxsize 3.0 truein \epsfbox{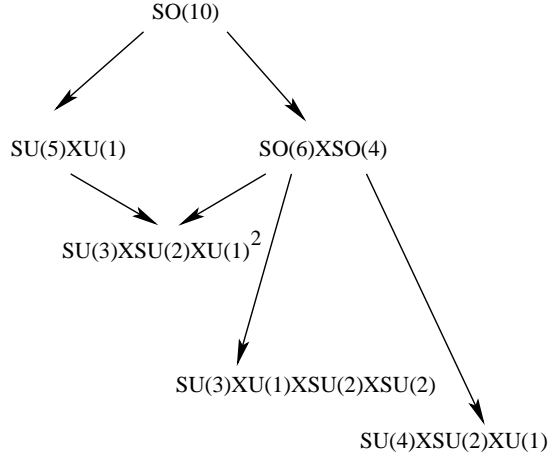}}
\caption{$SO(10)$ symmetry breaking patterns}  
\label{so10patterns}
\end{figure}

The analysis of this paper completes the classification of all
the $SO(10)$ subgroups in NAHE based models. The available symmetry
breaking patterns are depicted in figure \ref{so10patterns}.
This classification revealed that only the FSU5, PS, SLM, and LRS
subgroups produce viable spectra, whereas the
SU421 subgroup does not. The cases of LRS and SU421 models
also produced models that are completely free of gravitational
and gauge anomalies, whereas the other cases always contain
an anomalous $U(1)$. The origin of the differences between
the two cases can in fact be traced to the different
$N=4$ vacua from which the models emerge.

\section{Conclusions}\label{conclusion}

In this paper we studied the case of NAHE--based free fermionic
models with $SU(4)\times SU(2)\times U(1)$ $SO(10)$ subgroup.
This case offers some phenomenological advantages as compared to
some of the previous case studies, in the sense that it incorporates
the elegant string solution to
both the doublet--triplet and doublet--doublet splitting problems.
We demonstrated, however, that this choice of the $SO(10)$
subgroup as the unbroken four dimensional gauge group of a
free fermionic model cannot produce a realistic low energy spectrum.

Additionally, we discussed the important issue of the existence
of a anomalous $U(1)$ in the SU421 NAHE--based models. We showed
that models which are free of all Abelian and non--Abelian anomalies
exist in this class. The absence or presence of an anomalous
$U(1)$ is closely tied to the choice of the unbroken four
dimensional gauge group, and the ensued cancellation or non--cancellation
of the charges in specific orbits. We demonstrated that
the anomalous SU421 model, in contrast to the LRS model that was studied
in ref. \cite{ccf}, admits Abelian flat directions. This observation is in
agreement with the prevailing lore that superstring models generically
admit supersymmetric flat directions.

In conclusion we note that to date
the NAHE--based models represent the most realistic string models
constructed to date. This is particularly evident following the recent
neutrino observations that further support the embedding of the
Standard Model
spectrum in $SO(10)$ representations. As compared to related orbifold models
the NAHE--based models possess the distinctive advantage of admitting
three generations models with the $SO(10)$ embedding.
To advance further our understanding of the potential
relevance of string theory to experimental data it is
imperative that we delve further in the phenomenological
investigations of the most realistic case studies. In this paper
we demonstrated that in the NAHE--class a vacuum with three
generations and $SU(4)\times SU(2)\times U(1)$ $SO(10)$ subgroup is
not viable.

\section{Acknowledgments}

This work is supported in part by PPARC (AEF), and by Funds from the Pieter
Langerhuizen Lambertuszoon Trust held by the Royal
Holland Society of Sciences and Humanities and the VSB
Foundation (SN).

\vfill\eject

\appendix
{\textwidth=7.5in
\oddsidemargin=-18mm
\topmargin=-5mm
\renewcommand{\baselinestretch}{1.3}

\section{Anomaly-free model}

\noindent
\begin{tabular}{|c|c|c|rrrr|c||c|rr|}\hline
 SEC 				  &   $SU(4) \times$	  
& $Q_{R}$ & $Q_1$ & $Q_2$ & $Q_4$ & $Q_5$ &
$SU(2)_{3} \times$ & $SU(4)_H \times SU(2)_{H_3} \times$ & 
$Q_{7}$ & $Q_{8}$ \\
   				  &   $SU(2)$		  &   	  &   	  
&   	  &   	  &   	  &   $SU(2)_{6}$ &
$SU(2)_{H_1} \times SU(2)_{H_2}$	  &   	  &   	\\ \hline
  $\mb_1$ 			  &   $(\bar{4},1)$	  &   4	  &   -2  
&   0	  &   -2  &   0	  &
$(1,1)$	  &   $(1,1,1,1)$	  &   0	  &   0	\\
   				  &   $(\bar{4},1)$	  &   4	  &   2	  
&   0	  &   2	  &   0	  &
$(1,1)$	  &   $(1,1,1,1)$	  &   0	  &   0	\\
   				  &   $(\bar{4},1)$	  &   -4  &   -2  
&   0	  &   -2  &   0	  &
$(1,1)$	  &   $(1,1,1,1)$	  &   0	  &   0	\\
   				  &   $(\bar{4},1)$	  &   -4  &   2	  
&   0	  &   2	  &   0	  &
$(1,1)$	  &   $(1,1,1,1)$	  &   0	  &   0	\\\hline

  $\mb_2$ 			  &   $(\bar{4},1)$	  &   4	  &   0	  
&   -2  &   0	  &   2	  &
$(1,1)$	  &   $(1,1,1,1)$	  &   0	  &   0	\\
   				  &   $(\bar{4},1)$ 	  &   4	  &   0	  
&   2	  &   0	  &   -2  &
$(1,1)$	  &   $(1,1,1,1)$	  &   0	  &   0	\\
   				  &   $(\bar{4},1)$	  &   -4  &   0	  
&   -2  &   0	  &   2	  &
$(1,1)$	  &   $(1,1,1,1)$	  &   0	  &   0	\\
   				  &   $(\bar{4},1)$ 	  &   -4  &   0	  
&   2	  &   0	  &   -2  &
$(1,1)$	  &   $(1,1,1,1)$	  &   0	  &   0	\\\hline

  $\mb_3$ 			  &   $(4,2)$	  &   0	  &   0	  &   0	  
&   0	  &   0	  &
$(1,2)$	  &   $(1,1,1,1)$	  &   0	  &   0	\\\hline

  $S + \mb_1 + \mb_2 + $		  &   $(1,1)$	  &   0	  &   2  
&   -2	  &   0	  &   0
  &   $(1,1)$	  &   $(1,1,1,2)$	  &   0	  &   0	\\
  $\malpha + \mbeta$		  &   $(1,1)$	  &   0	  &   -2 &   2	  
&   0	  &   0	  &
 $(1,1)$	  &   $(1,1,2,1)$	  &   0	  &   0	\\
			 	  &   $(1,1)$	  &   0	  &   -2 &   2	  
&   0	  &   0	  &   $(1,1)$	  &
$(1,1,1,2)$	  &   0	  &   0	\\
				  &   $(1,1)$	  &   0	  &   2  &   -2   
&   0	  &   0	  &   $(1,1)$	  &
$(1,1,2,1)$	  &   0	  &   0	\\\hline

  $\mb_3 + \mbeta + 2\mgamma$	  &   $(1,1)$	  &   0	  &   0	  &   0	  
&   -2  &
  2	  &   $(2,1)$	  &   $(1,1,1,1)$	  &   4	  &   0	\\
   				  &   $(1,1)$	  &   0	  &   0	  &   0	  
&   2	  &   2	  &   $(1,2)$	  &
 $(1,1,1,1)$	  &   4	  &   0	\\
   				  &   $(1,1)$	  &   0	  &   0	  &   0	  
&   -2  &   -2  &   $(1,2)$	  &
 $(1,1,1,1)$	  &   4	  &   0	\\
   				  &   $(1,1)$	  &   0	  &   0	  &   0	  
&   2	  &   -2  &   $(2,1)$	  &
 $(1,1,1,1)$	  &   4	  &   0	\\
  				  &   $(1,1)$	  &   0	  &   0	  &   0	  
&   -2  &   2	  &   $(2,1)$	  &
$(1,1,1,1)$	  &   -4  &   0	\\
   				  &   $(1,1)$	  &   0	  &   0	  &   0	  
&   2	  &   2	  &   $(1,2)$	  &
 $(1,1,1,1)$	  &   -4  &   0	\\
   				  &   $(1,1)$	  &   0	  &   0	  &   0	  
&   -2  &   -2  &   $(1,2)$	  &
 $(1,1,1,1)$	  &   -4  &   0	\\
   				  &   $(1,1)$	  &   0	  &   0	  &   0	  
&   2	  &   -2  &   $(2,1)$	  &
 $(1,1,1,1)$	  &   -4  &   0	\\\hline

  $S + 2\mgamma$			  &   $(1,1)$	  &   4	  &   0	  
&   0	  &   0	  &   0	  &
$(1,1)$	  &   $(4,2,1,1)$	  &   0	  &   4	\\
  				  &   $(1,1)$	  &   -4  &   0	  &   0	  
&   0	  &   0	  &   $(1,1)$	  &
$(\bar{4},2,1,1)$	  &   0	  &   -4\\\hline

  $1 + S + \mb_3 +$	  &   $(1,2)$	  &   0	  &   2	  &   2	  &   0	  
&   0	  &
$(1,1)$	  &   $(1,1,2,1)$	  &   0	  &   0	\\
  $ \malpha + \mbeta + 2\mgamma$	  &   $(1,2)$	  &   0	  &   -2  
&   -2  &   0
 &   0	  &   $(1,1)$	  &   $(1,1,2,1)$	  &   0	  &   0	\\\hline

  $1 + \mb_1 + $	  &   $(4,1)$	  &   0	  &   0	  &   0	  &   0	  
&   0	  &
$(2,1)$	  &   $(1,1,1,2)$	  &   0	  &   0	\\
  $\mb_2 + 2\mgamma$		  &   		  &   	  &   	  &   	  
&   	  &   	  &   		  &
	 	  &   	  &   	\\\hline

  $1 + S + \mb_1 +$	  &   $(1,2)$	  &   0	  &   0	  &   0	  &   4	  
&   0	  &
$(1,1)$	  &   $(1,1,1,2)$	  &   0	  &   0	\\
  $\mb_2 + \mb_3 + 2\mgamma$		  &   $(1,2)$	  &   0	  &   0	  
&   0	  &   -4  &
  0	  &   $(1,1)$	  &   $(1,1,1,2)$	  &   0	  &   0	\\\hline
\end{tabular}

\vfill\eject
\section{Anomalous model}

\noindent
\begin{tabular}{|c|c|r|rrrrrr||c|rr|}\hline
 SEC	\& 	 	 &  $SU(4)$  & $Q_{R}$ & $Q_1$ & $Q_2$ & $Q_3$ & 
$Q_4$ & $Q_5$ &
$Q_6$ &  $SU(4)_H \times SU(2)_{H_3} \times$ 	  &  $Q_{7}$	 &  $Q_{8}$
	\\
Field 				 &$\times SO(5)$  &   	  &   	  &   	  &
&   	  &   	  &   	  &  $SU(2)_{H_2}$ 	 &   	 & 	\\
\hline
  $\bf 0$:                 &  &  &  &  &  &  &  &  & & & \\
  $\phi_1$, ${\bar\phi}_1$ & (1,1) &  0  &  0   & $\mp 1$ & $\mp 1$  &  0 
&  0
&  0 & $(1,1,1)$ & 0 & 0\\
  $\phi_2$, ${\bar\phi}_2$ & (1,1) &  0  &  0   & $\pm 1$ & $\mp 1$  &  0 
&  0
&  0 & $(1,1,1)$ & 0 & 0\\
  $\phi_3$, ${\bar\phi}_3$ & (1,1) &  0  &$\mp 1$&  0     & $\mp 1$  &  0 
&  0
&  0 & $(1,1,1)$ & 0 & 0\\
  $\phi_4$, ${\bar\phi}_4$ & (1,1) &  0  &$\pm 1$&  0     & $\mp 1$  &  0 
&  0
&  0 & $(1,1,1)$ & 0 & 0\\
  $\phi_5$, ${\bar\phi}_3$ & (1,1) &  0  &$\mp 1$& $\mp 1$&  0       &  0 
&  0
&  0 & $(1,1,1)$ & 0 & 0\\
  $\phi_6$, ${\bar\phi}_6$ & (1,1) &  0  &$\mp 1$& $\pm 1$&  0       &  0 
&  0
&  0 & $(1,1,1)$ & 0 & 0\\
\hline
  $\mb_1$ & $(\bar{4},1)$ &  -4 	  &  2 	  &  0 	  &  0 	  &  2 	  
&  0
&  0 	  &  $(1,1,1)$ 	 	 &  0 	 &  0	\\
   				 & $(\bar{4},1)$ &  4 	  &  2 	  &  0 	  
&  0 	  &  -2	  &  0 	  &  0
&  $(1,1,1)$ 	 	 &  0 	 &  0	\\
   				 & $(\bar{4},1)$ &  4 	  &  -2	  &  0 	  
&  0 	  &  2 	  &  0 	  &  0
&  $(1,1,1)$ 	 	 &  0 	 &  0	\\
 & $(\bar{4},1)$ &  -4 	  &  -2	  &  0 	  &  0 	  &  -2	  &  0 	  &  0
 &  $(1,1,1)$ 	 	 &  0 	 &  0	\\ \hline

  $\mb_2$ 			 & $(\bar{4},1)$ &  4 	  &  0 	  &  -2	  
&  0 	  &  0 	  &  -2	  &   0   &  $(1,1,1)$ 	  &  0 	 &  0	\\
 & $(\bar{4},1)$ &  -4 	  &  0 	  &  -2	  &  0 	  &  0 	  &  2 	  &  0
 &  $(1,1,1)$ 	  &  0 	 &  0	\\
& $(\bar{4},1)$ &  -4 	  &  0 	  &  2 	  &  0 	  &  0 	  &  -2	  &  0
 &  $(1,1,1)$ 	  &  0 	 &  0	\\
  & $(\bar{4},1)$ &  4 	  &  0 	  &  2 	  &  0 	  &  0 	  &  2 	  &  0
&  $(1,1,1)$ 	  &  0 	 &  0	\\\hline

  $\mb_3 \oplus \mb_3 + \zeta+ 2\gamma$	 &  $(4,4)$  	  &  0 	  &  0 	  
&  0 &  2 	  &  0 	  &  0 	  &  -2	  &  $(1,1,1)$ 	  &  0 	 &  0	\\
&  $(4,4)$ 	  &  0 	  &  0 	  &  0 	  &  2 	  &  0 	  &  0 	  &  2 	  &
 $(1,1,1)$ 	  &  0 	 &  0	\\\hline
$S + 2\mgamma$:			 &  $(1,1)$ 	  &  -4	  &  0 	  &  0 
&  0 	  &  0 	  &  0 &  0 	  &  $(6,1,1)$ 	  &  4	 &  0	\\
&  $(1,1)$ 	  &  4	  &  0 	  &  0 	  &  0 	  &  0 	  &  0 	  &  0 	 
& $(6,1,1)$ 	  &  -4	 &  0	\\
$S_1$, ${\bar S}_1$		 &  $(1,1)$ 	  & $\mp 4$	  &  0 	  
&  0 	  &  0 	  &  0 &  0 	  &  0 	  &  $(1,1,1)$ 	  & $\mp 4$ & 
$\pm 8$	\\
$S_2$, ${\bar S}_2$              &  $(1,1)$ 	  & $\mp 4$	  &  0 	  
& 0 	  &  0
  &  0 	  &  0 	  &  0 	  &
   				 $(1,1,1)$ 	  & $\mp 4$	 & $\mp 8$\\
\hline
  $\mb_2 + \malpha + 2\mgamma$&  &  &  &  &  &  &  &  & & & \\
 $S_3$, ${\bar S}_3$  				 &  $(1,1)$	  &  0 	  
&0 	  & $\mp 2$& 0 	  &$\mp
2$&$\mp 2$&$\mp 2$&  $(1,1,1)$ 	  &$\mp 4$&  0	\\
 $S_4$, ${\bar S}_4$  				 &  $(1,1)$	  &  0 	  
&0 	  & $\mp 2$& 0 	  &$\mp
2$&$\pm 2$&$\mp 2$&  $(1,1,1)$ 	  &$\pm 4$&  0	\\
  $S_5$, ${\bar S}_5$  				 &  $(1,1)$	  &  0 	  
&0 	  & $\mp 2$& 0 	  &$\mp
2$&$\mp 2$&$\pm 2$&  $(1,1,1)$ 	  &$\mp 4$&  0	\\
  $S_6$, ${\bar S}_6$ 				 &  $(1,1)$	  &  0 	  
&0 	  & $\mp 2$& 0 	  &$\mp
2$&$\pm 2$&$\pm 2$&  $(1,1,1)$ 	  &$\pm 4$&  0	\\
  $S_7$, ${\bar S}_7$  				 &  $(1,1)$	  &  0 	  
&0 	  & $\mp 2$& 0 	  &$\pm
2$&$\mp 2$&$\mp 2$&  $(1,1,1)$ 	  &$\pm 4$&  0	\\
  $S_8$, ${\bar S}_8$   	                 &  $(1,1)$ 	  &  0 	  
&0 	  & $\mp
2$& 0 	  &$\pm 2$&$\pm 2$&$\mp 2$&  $(1,1,1)$ 	  &$\mp 4$&  0	\\
  $S_9$, ${\bar S}_9$  				 &  $(1,1)$	  &  0 	  
&0 	  & $\mp 2$& 0 	  &$\pm
2$&$\mp 2$&$\pm 2$&  $(1,1,1)$ 	  &$\pm 4$&  0	\\
 $S_{10}$, ${\bar S}_{10}$			 &  $(1,1)$	  &  0 	  
&0 	  & $\mp 2$& 0 	  &$\pm
2$&$\pm 2$&$\pm 2$&  $(1,1,1)$ 	  &$\mp 4$&  0	\\
\hline
  $S + \mb_2 + \mb_3 + $		 &  $(1,1)$ 	  &  2 	  &  0 	  
&  2 	  &  2 	  &  0
 &  -2	  &  0 	  &  $(4,1,1)$ 	  &  0 	 &  0	\\
  $\malpha + \mbeta \pm \mgamma$	 &  $(1,1)$	  &  2 	  &  0 	  
&  -2	  &  2
&  0 	  &  2 	  &  0 	  &  $(4,1,1)$ 	  &  0 	 &  0	\\

   				 &  $(1,1)$	  &  2 	  &  0 	  &  -2	  
&  2 	  &  0 	  &  2 	  &  0 	  &
$(1,2,1)$ 	  &  -2	 &  0	\\
   				 &  $(1,1)$	  &  2 	  &  0 	  &  2 	  
&  2 	  &  0 	  &  -2	  &  0 	  &
$(1,2,1)$ 	  &  -2	 &  0	\\

  				 &  $(1,1)$ 	  &  -2	  &  0 	  &  -2	  
&  2 	  &  0 	  &  -2	  &  0 	  &
$(\bar{4},1,1)$ 	  &  0 	 &  0	\\
   				 &  $(1,1)$	  &  -2	  &  0 	  &  2 	  
&  2 	  &  0 	  &  2 	  &  0 	  &
$(\bar{4},1,1)$ 	  &  0 	 &  0	\\

   				 &  $(1,1)$	  &  -2	  &  0 	  &  2 	  
&  2 	  &  0 	  &  2 	  &  0 	  &
$(1,2,1)$ 	  &  2 	 &  0	\\
   				 &  $(1,1)$	  &  -2	  &  0 	  &  -2	  
&  2 	  &  0 	  &  -2	  &  0 	  &
$(1,2,1)$ 	  &  2 	 &  0	\\ \hline

\end{tabular}
\vfill\eject

\noindent
\begin{tabular}{|c|c|r|rrrrrr||c|rr|}\hline
 SEC				 &  $SU(4) \times SO(5)$ 	  & $Q_{R}$ 
& $Q_1$ & $Q_2$ & $Q_3$ & $Q_4$ &
$Q_5$ & $Q_6$ &  $SU(4)_H \times SU(2)_{H_3} \times$ 	  &  $Q_{7}$	 
&  $Q_{8}$
\\
   				 &   	  &   	  &   	  &   	  &   	  
&   	  &   	  &   	  &
$SU(2)_{H_2}$ 	  &   	 & 	\\\hline

  $S + \mb_1 + \mb_3 + $		 &  $(1,1)$ 	  &  2 	  &  -2   
&  0 	  &  2 	  &  -2
 & 0	&  0 	  &  $(4,1,1)$ 	  &  0 	 &  0	\\
  $\malpha + \mbeta \pm \mgamma$	 &  $(1,1)$	  &  2 	  &  2 	  
&  0 	  &  2
&  2 	  & 0	&  0 	  &  $(4,1,1)$ 	  &  0 	 &  0	\\

   				 &  $(1,1)$	  &  2 	  &  2    &  0	  
&  2 	  &  2 	  & 0	&  0 	  &
$(1,2,1)$ 	  &  -2	 &  0	\\
   				 &  $(1,1)$	  &  2 	  &  -2   &  0	  
&  2 	  &  -2	  & 0	&  0 	  &
$(1,2,1)$ 	  &  -2	 &  0	\\

  				 &  $(1,1)$ 	  &  -2	  &  2 	  &  0 	  
&  2 	  &  -2	  & 0	&  0 	  &
$(\bar{4},1,1)$ 	  &  0 	 &  0	\\
   				 &  $(1,1)$	  &  -2	  &  -2   &  0 	  
&  2 	  &  2 	  & 0	&  0 	  &
$(\bar{4},1,1)$ 	  &  0 	 &  0	\\

   				 &  $(1,1)$	  &  -2	  &  -2   &  0 	  
&  2 	  &  2 	  & 0	&  0 	  &
$(1,2,1)$ 	  &  2 	 &  0	\\
   				 &  $(1,1)$	  &  -2	  &  2    &  0	  
&  2 	  &  -2	  & 0	&  0 	  &
$(1,2,1)$ 	  &  2 	 &  0	\\\hline

  $S + \mb_1 + \mb_2 + $		 &  $(1,4)$ 	  &  0 	  &  -2   
&  -2	  &  0 	  &  0
 &  0 	  &  0 	  &  $(1,1,2)$ 	  &  0 	 &  0	\\
  $\malpha + \mbeta \ \oplus $	 &  	 	  &   	  &   	  &   	  
&   	  &   	  &
  &   	  &  	 	  &   	 &  	\\
  $S + \mb_1 +\mb_2 + $		 &  $(1,4)$	  &  0 	  &  2	  
&  2	  &  0 	  &
0 	  &  0 	  &  0	  &  $(1,1,2)$ 	  &  0 	 &  0	\\
  $\malpha + \mbeta+\zeta+2\gamma$	 &  	 	  &   	  &   	  &   	  
&   	  &   	  &
&   	  &  	 	  &   	 &  	\\ \hline

  $1 + b_1 + $	 &  $(1,1)$ 	  &  2 	  &  0 	  &  2	  & 2	  &  0 	  
&  -2	  &  0
	  &  $(1,1,2)$ 	  &  0 	 &  0	\\
  $\malpha + \mbeta \pm \mgamma$	 &  $(1,1)$	  &  2 	  &  0 	  
&  -2	  & 2	  &
0 	  &  2 	  &  0 	  &  $(1,1,2)$ 	  &  0 	 &  0	\\
   				 &  $(1,1)$	  &  -2	  &  0 	  &  -2	  
& 2	  &  0 	  &  -2	  &  0 	  &
$(1,1,2)$ 	  &  0 	 &  0	\\
   				 &  $(1,1)$	  &  -2	  &  0 	  &  2	  
& 2	  &  0 	  &  2 	  &  0 	  &
$(1,1,2)$ 	  &  0 	 &  0	\\ \hline

  $1 + b_2 + $	 &  $(1,1)$ 	  &  2 	  &  2	  &  0 	  & 2	  &  2 	  
&  0 	  &  0
	  &  $(1,1,2)$ 	  &  0 	 &  0	\\
  $\malpha + \mbeta \pm \mgamma$	 &  $(1,1)$	  &  2 	  &  -2	  
&  0 	  & 2	  &
-2	  &  0 	  &  0 	  &  $(1,1,2)$ 	  &  0 	 &  0	\\
   				 &  $(1,1)$	  &  -2	  &  -2	  &  0 	  
& 2	  &  2 	  &  0 	  &  0 	  &
$(1,1,2)$ 	  &  0 	 &  0	\\
   				 &  $(1,1)$	  &  -2	  &  2	  &  0 	  
& 2	  &  -2	  &  0 	  &  0 	  &
$(1,1,2)$ 	  &  0 	 &  0	\\\hline
\end{tabular}
\hfill\vfill\eject}

\section{D-Flat Basis for Anomalous SU421 Model}

\noindent
\begin{flushleft}
\begin{tabular}{|l|l|rrrrrr|rrrrrrrrrr|}\hline
N&$Q_A$& $\phi$ & & & & & &$S$& & & & & & & & & \\
 &     &    1   &2&3&4&5&6&1  &2&3&4&5&6&7&8&9&10\\
\hline
$D_1$&-8& 2& 0& 0& 0& 0& 0& 0& 0& 0& 0&-1&-1& 0&-2&-1& 1\\
$D_2$&-8& 0& 2& 0& 0& 0& 0& 0& 0& 0& 0& 1& 1& 0& 2& 1&-1\\
$D_3$&-8& 0& 0& 2& 0& 0&-2& 0& 0& 0& 0&-1&-1& 0&-2&-1& 1\\
$D_4$&-8& 0& 0& 0& 2& 0& 2& 0& 0& 0& 0& 1& 1& 0& 2& 1&-1\\
$D_5$& 0& 0& 0& 0& 0& 1&-1& 0& 0& 0& 0&-1&-1& 0&-2&-1& 1\\
$D_6$& 0& 0& 0& 0& 0& 0& 0& 0& 0& 1& 0&-1& 0& 0&-1& 0& 1\\
$D_7$& 0& 0& 0& 0& 0& 0& 0& 0& 0& 0& 1& 0&-1& 0&-1& 0& 1\\
$D_8$& 0& 0& 0& 0& 0& 0& 0& 0& 0& 0& 0& 0& 0& 1&-1&-1& 1\\
\hline
\end{tabular}
\end{flushleft}
\noindent The 8 elements of a basis set of VEVs for forming $D$-flat
directions for the anomalous SU421 model. First column entries give
basis element designations. The second column specifies the
anomalous charge of each element. The remaining columns specify the ratio of
the norms of the VEVs of the scalar fields. All 16 fields are part of
vector-like pairs of scalars and a negative norm indicates the
vector partner field acquires the VEVs. For each of the $D_i$ in the table
there are corresponding ${\bar D}_i$ with sets of norms of VEVs of opposite
signs.
\hfill\vfill\eject
\newpage



\vfill\eject

{{\oddsidemargin  10.5pt \evensidemargin  10.5pt
\textheight  612pt \textwidth  432pt
\headheight  12pt \headsep  20pt
\footheight  12pt \footskip  40pt

\bibliographystyle{unsrt}

\begin{thebibliography}{99}

\bibitem{cfn1}{G.B. Cleaver \etal. \PLB{445}{1999}{135};
				   	\IJMP{16}{2001}{425};
					\NPB{593}{2001}{471};
					\MODA{15}{2000}{1191};
					\IJMP{16}{2001}{3565};
					\NPB{620}{2002}{259}.}

\bibitem{fsu5}{I. Antoniadis, J. Ellis, J. Hagelin, and D.V. Nanopoulos,
				\PLB{231}{1989}{65};\\
	J.L. Lopez, D.V. Nanopoulos and K. Yuan, \NPB{399}{1993}{654}.}

\bibitem{fny} A.E. Faraggi, D.V. Nanopoulos, and K. Yuan,
              \NPB{335}{1990}{347}; \\
	A.E. Faraggi, \PRD{46}{1992}{3204};

\bibitem{slm}{\AEF, \PLB{278}{1992}{131}; \NPB{387}{1992}{239}.}

\bibitem{alr}{I. Antoniadis, G.K. Leontaris and J. Rizos,
    			\PLB{245}{1990}{161};\\
              G.K. Leontaris and J. Rizos, \NPB{554}{1999}{3}.}

\bibitem{cfs} G.B. Cleaver, A.E. Faraggi and C. Savage,
					\PRD{63}{2001}{066001}.


\bibitem{ps} \AEF, \NPB{428}{1994}{111}; \PLB{520}{2001}{337}.

\bibitem{otherdts} E. Witten,  \NPB{258}{1985}{75};\\
J.D. Breit, B.A. Ovrut and G.C. Segre, \PLB{158}{1985}{33};\\
L.E. Ibanez, J.E. Kim, H.P. Nilles and F. Quevedo, \PLB{191}{1987}{202}.
\bibitem{fsu5dts} I. Antoniadis, J. Ellis, J. Hagelin and D.V. Nanopoulos,
\PLB{194}{1987}{231}.

\bibitem{yukawa} \AEF, \PLB{274}{1992}{47}; \PRD{47}{1993}{5021};
			\NPB{403}{1993}{101}; \NPB{407}{1993}{57}.


\bibitem{ccf} G.B. Cleaver, D.J. Clements and A.E. Faraggi,
					\PRD{65}{2002}{106003}.

\bibitem{nahe}{A.E. Faraggi and D.V. Nanopoulos, \PRD{48}{1993}{3288}.}

\bibitem{lrsfcnc} M. Pospelov, \PRD{59}{1997}{259}, and references therein.
\bibitem{bidoublet}{R.N. Mohapatra and A. Rasin, \PRD{54}{1996}{5835};\\
C.S. Aulakh, A. Melfo, A. Rasin and G. Senjanovic, \PRD{58}{1998}{115007}.}
\bibitem{fff}{I. Antoniadis, C. Bachas, and C. Kounnas, \NPB{289}{1987}{87};\\
               H. Kawai, D.C. Lewellen, and S.H.-H. Tye, \NPB{288}{1987}{1}.}
\bibitem{kn}{T. Kobayashi and H. Nakano, \NPB{496}{1997}{103}.}
\bibitem{cf1}{G.B. Cleaver and A.E. Faraggi,
             \IJMP{14}{1999}{2335}.}

\bibitem{custodial} \AEF, \PLB{339}{1994}{223}.

\bibitem{foc}\AEF, \PLB{326}{1994}{62}; \PLB{544}{2002}{207};\\
	 J. Ellis, \AEF~and D.V. Nanopoulos, \PLB{419}{1998}{123};\\
	 P. Berglund \etal, \PLB{433}{1998}{269}; \IJMP{15}{2000}{1345}.


\bibitem{dsw} M. Dine, N. Seiberg and E. Witten, \NPB{289}{1987}{589};\\
		J. Atick, L. Dixon and A. Sen, \NPB{292}{1987}{109}.


\end{thebibliography}
}}
\vfill\eject

\end{document}